\documentclass{ws-mpla}

\RequirePackage{xspace}

\usepackage{relsize}
\def\babar{\mbox{\slshape B\kern-0.1em{\smaller A}\kern-0.1em
    B\kern-0.1em{\smaller A\kern-0.2em R}}}

\def\cp    {\ensuremath{C\kern-0.2em P}}
\def\cpv   {\ensuremath{C\kern-0.2em P\kern-1.0em /}}

\def\vub   {\ensuremath{|V_{ub}|}}
\def\vcb   {\ensuremath{|V_{cb}|}}

\def\deltam{\ensuremath{\delta m}}
\def\TBY   {\ensuremath{\theta_{\Bz, D^*\ell}}}

\def\de    {\ensuremath{\Delta E}}
\def\mes   {\ensuremath{m_{ES}}}

\def\lbar{\ensuremath{\overline{\Lambda}}}
\def\lone{\ensuremath{\lambda_1}}
\def\ltwo{\ensuremath{\lambda_2}}

\def\cbf {\ensuremath{{\cal B}}}
\def\clu {\ensuremath{{\cal L}}}

\def\mmiss{\ensuremath{{m_{miss}^2}}}
\def\rusl{\ensuremath{{R_{u}}}}

\def\mX{\ensuremath{{m_X}}}
\def\mx{\ensuremath{{m_X}}}
\def\mxcut{\ensuremath{{m_X^{cut}}}}

\def\pt{\ensuremath{{p_\perp}}}

\def\Bpilnu    {\ensuremath{\Bb\to \pi\ell\nub}}
\def\Bmpilnu   {\ensuremath{\Bm\to \pi^0\ell^-\nub}}

\def\Brholnu   {\ensuremath{\Bb\to \rho\ell\nub}}

\def\Brhoenu   {\ensuremath{\Bb\to \rho e\nub}}
\def\Bzrhoenu  {\ensuremath{\Bz\to \rho^- e^+\nub}}

\def\Bdlnu     {\ensuremath{\Bb\to D\ell\nub}}
\def\Bdstarlnu {\ensuremath{\Bb\to \Dstar \ell \nub}}
\def\Bzdstarlnu {\ensuremath{\Bzb\to \Dstarp \ell^- \nub}}
\def\Bzdstarenu {\ensuremath{\Bzb\to \Dstarp e^- \nub}}

\def\bsg     {\ensuremath{b\to s\gamma}}

\def\Bxenu   {\ensuremath{\Bb\to Xe^-\nu}}

\newcommand {\Bxlnu}{\ensuremath{\Bb \rightarrow X \ell \bar{\nu}}}
\newcommand {\Bxclnu}{\ensuremath{\Bb \rightarrow X_c \ell \bar{\nu}}}
\newcommand {\Bxulnu}{\ensuremath{\Bb \rightarrow X_u \ell \bar{\nu}}}

\def\Bxulnu  {\ensuremath{\Bb\to X_{u}\ell\nub}}

\def\Bxclnu  {\ensuremath{\Bb\to X_{c}\ell\nub}}

\def\breco   {\ensuremath{B_{reco}}}

\def\eg   {{\it e.g.}}

\def\de        {\ensuremath {\Delta E^{*}}}

\def\nub        {\ensuremath{\bar{\nu}}\xspace}

\def\qqbar {\ensuremath{q\overline q}\xspace}

\def\piz   {\ensuremath{\pi^0}\xspace}

\def\pip   {\ensuremath{\pi^+}\xspace}
\def\pim   {\ensuremath{\pi^-}\xspace}

\def\Kbar  {\kern 0.2em\overline{\kern -0.2em K}{}\xspace}

\def\Kz    {\ensuremath{K^0}\xspace}
\def\Kzb   {\ensuremath{\Kbar^0}\xspace}
\def\KzKzb {\ensuremath{\Kz \kern -0.16em \Kzb}\xspace}
\def\Kp    {\ensuremath{K^+}\xspace}
\def\Km    {\ensuremath{K^-}\xspace}

\def\KpKm  {\ensuremath{\Kp \kern -0.16em \Km}\xspace}
\def\KS    {\ensuremath{K^0_{\scriptscriptstyle S}}\xspace}

\def\Dbar    {\kern 0.2em\bar{\kern -0.2em D}{}\xspace}
\def\Db      {\ensuremath{\Dbar}\xspace}
\def\Dz      {\ensuremath{D^0}\xspace}
\def\Dzb     {\ensuremath{\Dbar^0}\xspace}
\def\DzDzb   {\ensuremath{\Dz {\kern -0.16em \Dzb}}\xspace}
\def\Dp      {\ensuremath{D^+}\xspace}
\def\Dm      {\ensuremath{D^-}\xspace}

\def\DpDm    {\ensuremath{\Dp {\kern -0.16em \Dm}}\xspace}
\def\Dstar   {\ensuremath{D^*}\xspace}

\def\Dstarp  {\ensuremath{D^{*+}}\xspace}

\def\B       {\ensuremath{B}\xspace}
\def\Bbar    {\kern 0.18em\bar{\kern -0.18em B}{}\xspace}
\def\Bb      {\ensuremath{\Bbar}\xspace}
\def\BB      {\ensuremath{B\Bbar}\xspace} 
\def\Bz      {\ensuremath{B^0}\xspace}
\def\Bzb     {\ensuremath{\Bbar^0}\xspace}
\def\BzBzb   {\ensuremath{\Bz {\kern -0.16em \Bzb}}\xspace}
\def\Bu      {\ensuremath{B^+}\xspace}
\def\Bub     {\ensuremath{B^-}\xspace}
\def\Bp      {\ensuremath{\Bu}\xspace}
\def\Bm      {\ensuremath{\Bub}\xspace}

\def\BpBm    {\ensuremath{\Bu {\kern -0.16em \Bub}}\xspace}

\mathchardef\Upsilon="7107
\def\Y#1S{\ensuremath{\Upsilon{(#1S)}}\xspace}

\def\FourS {\Y4S}

\mathchardef\Deltares="7101
\mathchardef\Xi="7104
\mathchardef\Lambda="7103
\mathchardef\Sigma="7106
\mathchardef\Omega="710A

\def\Deltabar{\kern 0.25em\overline{\kern -0.25em \Deltares}{}\xspace}
\def\Lbar{\kern 0.2em\overline{\kern -0.2em\Lambda\kern 0.05em}\kern-0.05em{}\xspace}
\def\Sigbar{\kern 0.2em\overline{\kern -0.2em \Sigma}{}\xspace}
\def\Xibar{\kern 0.2em\overline{\kern -0.2em \Xi}{}\xspace}
\def\Obar{\kern 0.2em\overline{\kern -0.2em \Omega}{}\xspace}
\def\Nbar{\kern 0.2em\overline{\kern -0.2em N}{}\xspace}
\def\Xb{\kern 0.2em\overline{\kern -0.2em X}{}\xspace}

\def\BR         {{\ensuremath{\cal B}\xspace}}

\def\pt         {\mbox{$p_\perp$}\xspace}
\def\mes        {\mbox{$m_{\rm ES}$}\xspace}

\newcommand{\tev}{\ensuremath{\mathrm{\,Te\kern -0.1em V}}\xspace}
\newcommand{\gev}{\ensuremath{\mathrm{\,Ge\kern -0.1em V}}\xspace}
\newcommand{\mev}{\ensuremath{\mathrm{\,Me\kern -0.1em V}}\xspace}
\newcommand{\kev}{\ensuremath{\mathrm{\,ke\kern -0.1em V}}\xspace}
\newcommand{\ev}{\ensuremath{\mathrm{\,e\kern -0.1em V}}\xspace}
\newcommand{\gevc}{\ensuremath{{\mathrm{\,Ge\kern -0.1em V\!/}c}}\xspace}
\newcommand{\mevc}{\ensuremath{{\mathrm{\,Me\kern -0.1em V\!/}c}}\xspace}
\newcommand{\gevcc}{\ensuremath{{\mathrm{\,Ge\kern -0.1em V\!/}c^2}}\xspace}
\newcommand{\mevcc}{\ensuremath{{\mathrm{\,Me\kern -0.1em V\!/}c^2}}\xspace}

\def\nb   {\ensuremath{\rm \,nb}\xspace}

\def\invfb     {\ensuremath{\mbox{\,fb}^{-1}}\xspace}

\def\to                 {\ensuremath{\rightarrow}\xspace}

\def\pep2{PEP-II}

\def\gsim{{~\raise.15em\hbox{$>$}\kern-.85em
          \lower.35em\hbox{$\sim$}~}\xspace}
\def\lsim{{~\raise.15em\hbox{$<$}\kern-.85em
          \lower.35em\hbox{$\sim$}~}\xspace}

\newcommand{\MSb}{\ensuremath{\overline{\mathrm{MS}}}\xspace}

\def\Vub  {\ensuremath{|V_{ub}|}\xspace}

\xspace

\newcommand{\lqcd}{\ensuremath{\Lambda_{\mathrm{QCD}}}\xspace}

\def\jetset74   {\mbox{\tt Jetset \hspace{-0.5em}7.\hspace{-0.2em}4}\xspace}

\def\allepsu{0.342}
\def\allepsmx{0.733} 

\def\allbrbrVal{2.06}
\def\allbrbrEcstat{0.25}
\def\allbrbrEsyst{0.23}
\def\allbrbrEthHi{0.36}

\def\allbr{ 2.24}
\def\allbrE{ 0.27}
\def\allbrEsyst{ 0.26}
\def\allbrEthHi{ 0.39}

\def\allvub{ 4.62}
\def\allvubE{ 0.28}
\def\allvubEsyst{ 0.27}
\def\allvubEthHi{ 0.40}
\def\allvubEtheo{ 0.26}

\begin{document}

\markboth{Urs Langenegger}
{Semileptonic $B$ Decays at \babar}

%
%

\title{Semileptonic $B$ Decays  at \babar\footnote{Originally based on
    an invited seminar first given at CERN, with updated results.}  }

\author{\footnotesize Urs Langenegger\footnote{Work supported in part by Department of Energy contract
  DE-AC03-76SF00515.}}

\address{Physikalisches Institut, Philosophenweg 12, D-69120
  Heidelberg, Germany\footnote{Address before October 2003: Stanford
  Linear Accelerator Center, Stanford, CA 94309, USA}
\\
ursl@physi.uni-heidelberg.de}

\maketitle


\begin{abstract}
 
 We  present results  on  semileptonic $B$  decays  obtained with  the
 \babar\  detector.  The  large  data set  accumulated  at the  PEP-II
 asymmetric-energy  $B$-Factory allows  a  new measurement  technique,
 where the hadronic decay of  one $B$ meson is fully reconstructed and
 the  semileptonic  decay of  the  recoiling  $B$  meson is  studied.  
 Traditional analysis  techniques of inclusive and  exclusive $B$ decays
 complement  this approach  with very  high statistics  data  samples. 
 These  measurements  play an  important  role  in  the tests  of  the
 description of \cp\ violation in the Standard Model: The determinations
 of  the  Cabibbo-Kobayashi-Maskawa matrix  elements  \vcb\ and  \vub\ 
 provide  constraints   on  the  unitarity   of  the  CKM   triangle.  
 Furthermore,  the  experimental measurement  of  parameters of  Heavy
 Quark  Effective  Theory  test  the consistency  of  the  theoretical
 description of semileptonic $B$ decays.

\end{abstract}

\ccode{PACS Nos.: 13.20.He Decays of beauty mesons and
      12.15.Nh Determination of Kobayashi-Maskawa matrix elements}


\section{Introduction}    
The  principal  motivation  for  the  study of  flavor  physics  is  a
comprehensive  test   of  the  Standard  Model   description  of  \cp\
violation.  Semileptonic  $B$ decays allow for  a direct determination
of \vcb\ and \vub, two elements of the Cabibbo-Kobayashi-Maskawa (CKM)
quark mixing matrix. In the unitarity triangle, the precision of \vcb\
affects  constraints   derived  from  kaon  decays   and  the  overall
normalization, while  the uncertainty in \vub\ dominates  the error of
the length of the side opposite  the angle $\beta$.  As this angle can
be  measured  very cleanly  in  time-dependent  \cp\ asymmetries,  the
errors of \vub\ must be  model independent, well understood, and small
before any discrepancies between sides and angles could be interpreted
as  new physics.  This  is not  yet the  case: the  error in  \vub\ is
larger     than     $10\%$     and    dominated     by     theoretical
uncertainties.\cite{Ligeti:2003hp,Luke:2003nu}

In the  theoretical description of semileptonic $B$  decays, the large
mass  of  the $b$  quark  plays a  central  role  by implying  special
symmetries and a hard  scale.\cite{intros} This is formulated by Heavy
Quark Effective  Theory\cite{HQET} (HQET) for exclusive  decays and an
Operator Product Expansion\cite{ope} (OPE) for inclusive decays.  Both
provide  systematic expansions  of  the (differential)  decay rate  in
terms   of  $\lqcd/m_b$   and   $\alpha_s(m_b)$.   Here   incalculable
quantities are parametrized in terms of expectation values of hadronic
matrix  elements, which  can  be  related to  the  shape (moments)  of
inclusive  decay spectra.   The large  rate of  Cabibbo-favored decays
\Bxclnu\ allows for precise measurements of the relevant distributions
and  the determination  of HQET  parameters.  This  provides precision
determinations  of  \vcb\  and  stringent quantitative  tests  of  the
consistency of the theory.

The situation is different for Cabibbo-suppressed \Bxulnu\ decays: the
large rate for \Bxclnu\ decays  constitutes a background that is about
50   times  larger,   overlapping  in   most  of   the   phase  space.
Experimentally,  selection   criteria  are  applied   to  reduce  this
background, but can lead to problems in the theoretical description.

The experimental approach to semileptonic decays can be separated into
two classes: Exclusive decays  reconstruct one signal decay mode, \eg,
\Bzdstarlnu.  Even  though the  neutrino cannot be  measured directly,
this approach  is relatively straightforward.   The inclusive analysis
of  semileptonic decays  is  often  based on  the  measurement of  the
charged  lepton  alone, or  by  combining  hadronic  final states  $X$
without disentangling specific resonances.



\section{The \babar\ Detector}    
The measurements  presented here  are based on  data collected  by the
\babar\ detector\cite{Aubert:2001tu} at  the PEP-II asymmetric energy
$e^+e^-$ collider  near the \FourS\  resonance.  Most of  the analyses
use  an integrated  luminosity of  about $80\invfb$,  corresponding to
about  89 million  \BB\ pairs.   The \FourS\  resonance is  just above
threshold for  the decay into a  pair of $B$ mesons  (either \BpBm\ or
\BzBzb), without any  other fragmentation particles.  Furthermore, the
two $B$ mesons have a low  and known momentum of $p_B^* = 320\mevc$ in
the center-of-mass system (CMS)\footnote{All variables measured in the
CMS  frame,  \eg,  $p_B^*$,  are  marked with  a  star.},  leading  to
spherically  symmetric  decays.   This  is  different  for  $q\bar{q}$
continuum  processes  (where $q  =  u,d,s,c$),  which  exhibit a  more
jet-like structure.  This is  exploited with event shape variables and
neural networks.

At  a CMS  energy  of $\sqrt{s}=  10.58\gev$,  the \FourS\  production
cross-section amounts  to about $1.1\nb$.  This corresponds  to a rate
of about $10\, \BB$  pairs$/\mbox{sec}$ at an instantaneous luminosity
of $10^{34}\,\mbox{cm}^{-2}\,\mbox{s}^{-1}$.  Given the total hadronic
continuum  cross section  of ca.   $3.5\nb$, the  resulting  signal to
background  ratio is  much  higher than  at  hadronic colliders.   The
background  from  continuum   processes  is  determined  in  dedicated
``off-peak''  runs, where  the CMS  energy is  lowered to  $\sqrt{s} =
10.54\gev$.

A five-layer  silicon vertex tracker provides  precision vertexing and
low-momentum  charged particle  tracking, down  to  transverse momenta
$\pt \sim 50\mev$. This is especially important for the reconstruction
of $\Dstarp\to \Dz\pip_s$ decays,  where the ``slow'' pion $\pi_s$ has
very  low  energies.  A  40-layer  driftchamber  surrounds the  vertex
detector and  complements the momentum measurement.   In addition, the
$dE/dx$ measurements  are used  in the identification  of low-momentum
electrons.   The  DIRC  provides  detection  of  internally  reflected
Cherenkov   light  used   in  charged   hadron   identification.   The
electromagnetic CsI(Tl) crystal calorimeter is the most important detector
for  electron identification  (by  means  of the  ratio  $E/p$ of  the
deposited  energy  $E$  and  the  momentum  $p$).   In  addition,  its
measurements  of  neutral  particles  is  crucial  for  the  inclusive
determination  of the  invariant  mass \mx\  in  \Bxlnu\ decays.   The
detector is surrounded by a superconducting coil (providing a magnetic
field of  $1.5\,\hbox{T}$) and its  instrumented flux return,  used in
the identification of muons.

The boosted CMS at \babar\ leads to a limited coverage of about $85\%$
of the solid angle in the  CMS. This is a notable disadvantage for the
reconstruction  of  neutrinos  from  the missing  momentum,  as  about
$1\gev$ of energy is missed per event (on average).

\section{Recoil Physics}    
The very  large luminosity  at the \babar\  detector allows for  a new
paradigm  for  the  systematic  study  of  semileptonic  $B$  decays.  
Traditionally\cite{Albrecht:1993pu}  events are  selected (``tagged'')
by a high-momentum lepton, signaling  the semileptonic decay of one of
the $B$ mesons and thereby reducing \qqbar\ continuum events.

At \babar, an alternative  event tagging technique has been developed:
the hadronic  decay of one  $B$ meson (\breco) is  fully reconstructed
and the semileptonic decay of the other $B$ meson is identified by the
presence  of an  electron or  muon.  This  approach results  in  a low
overall event  selection efficiency, but allows  for the determination
of  the  momentum, charge,  and  flavor of  the  $B$  mesons. It  also
provides a direct determination of the hadronic final state in \Bxlnu\ 
decays,  as all  particles  in  the recoil  of  the \breco\  candidate
originate  from the  other $B$  meson decaying  semileptonically. This
method  also offers  a  promising way  to  study semileptonic  $\Bb\to
X\tau\bar{\nu}_\tau$ decays.

\begin{figure}[!ht]
 \begin{center}
  \unitlength1.0cm 
  \begin{picture}(25.,6.)
   \put(  0.0,  0.0){\includegraphics[width=0.49\textwidth]{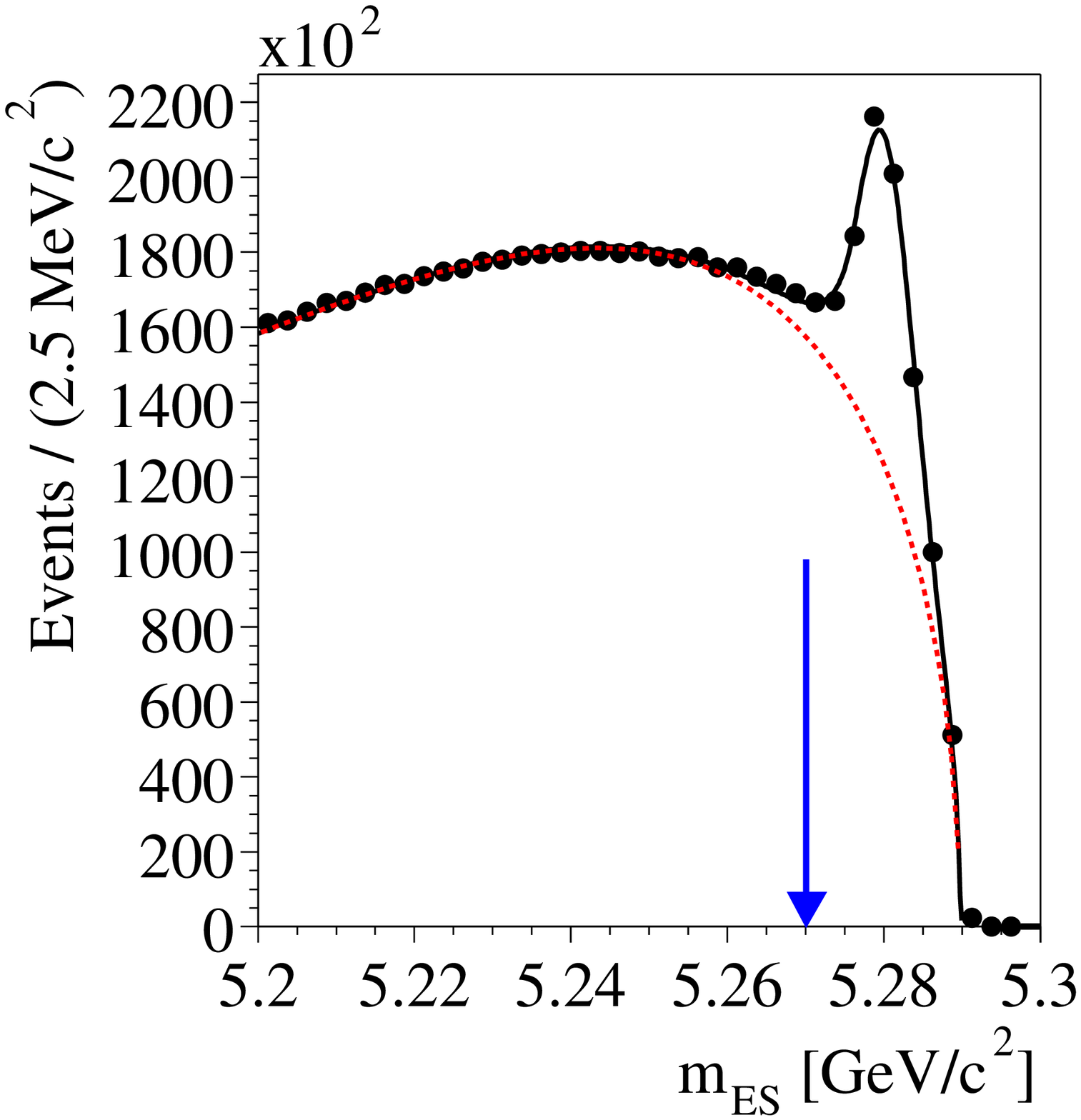}}
   \put(  6.8,  0.0){\includegraphics[width=0.49\textwidth]{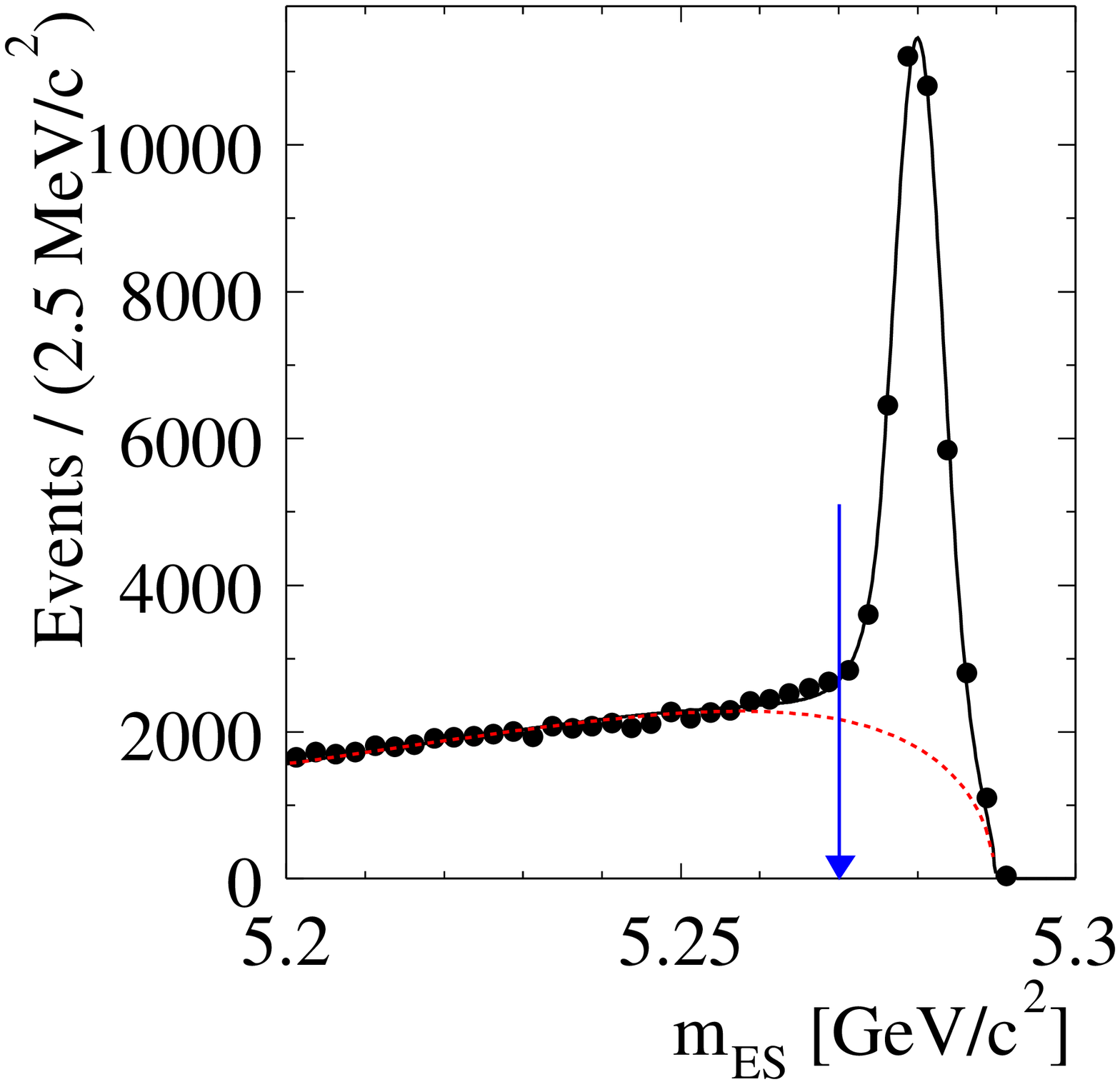}}
   \put(  1.7,  5.0){{\large\bf a)}}
   \put(  8.5,  5.0){{\large\bf b)}}
  \end{picture}
  \caption{The \mes\ distributions for fully reconstructed hadronic
    $B$  decays  used  in  the   event  selection  for  the  study  of
    semileptonic $B$ decays. a) With no requirement on the rest of the
    event,  the purity  amounts  to 26\%.  b)  With a  $p^* >  1\gevc$
    lepton,  the purity  improves  to 67\%.  The  arrows indicate  the
    minimum \mes\ requirement used in  the selection of signal events. 
  }
  \label{f:breco}
 \end{center}
\end{figure}

A  very large  sample  of  $B$ mesons  is  reconstructed by  selecting
hadronic decays\footnote{Charge conjugation  is implied throughout the
  text.} $\breco\to\Db Y^{+}, \Db^*  Y^{+}$, where the hadronic system
$Y^+$ consists of $n_1\pi^{\pm}\, n_2K^{\pm}\, n_3\KS\, n_4\piz$, with
$n_1 +  n_2 \leq  5$, $n_3 \leq  2$, and  $n_4 \leq 2$.  The kinematic
consistency  of  $B_{reco}$  candidates   is  checked  with  the  beam
energy-substituted mass $\mes =  \sqrt{s/4 - \vec{p}^{\,2}_B}$ and the
energy difference $\Delta  E = E_B - \sqrt{s}/2$,  where $\sqrt{s}$ is
the  total   energy  and  $(E_B,  \vec{p}_B)$   denotes  the  momentum
four-vector of the  $B_{reco}$ candidate in the CMS.   For each of the
reconstructed $B$ decay  modes, the purity ${\cal P}$  is estimated as
the   signal  fraction   in  events   with  \mes$>   5.27$\gevcc  (see
Fig.~\ref{f:breco}).  A priori, the purity  of this sample is low, but
improves substantially  in conjunction with the requirement  of a high
momentum lepton in  the recoil.  By combining more  than 300 modes, at
least one \breco\  candidate is reconstructed in 0.3\%  (0.5\%) of the
\BzBzb\ (\BpBm)  events.  In events  with more than  one reconstructed
\breco\ candidate, we select the decay mode with the highest purity.

\section{\Bzdstarlnu}
While the decay mode \Bzdstarlnu\  has a large branching fraction, the
measurements  so far  are not  very consistent---recent  results range
from   $(4.59\pm0.46)\%$  to   $(6.09\pm0.44)\%$.    The  very   large
luminosity  at  \babar\  opens  new possibilities  for  the  precision
determination of this decay.

The  theoretical description  of $B\to  D^{(*)} \ell  \nub$  decays in
terms  of  HQET\cite{HQET}   predicts  the  differential  decay  rate
schematically as

\begin{equation}
 \frac{d\Gamma(\Bb\to D^{(*)}\ell\nub)}{dw} = 
 {\cal K}\cdot \vcb^2\cdot
 \left\{%
  \begin{array}{ll}
   (w^2 - 1)^{1/2}\cdot {\cal F}_*^2(w) &\\
   (w^2 - 1)^{3/2}\cdot {\cal F}^2(w)   &
  \end{array}
  \right.%
\label{e:diff}
\end{equation}

\noindent where $w \equiv v_B\cdot v_{D^*} =
E_{D^*}/m_{D^*}$, ${\cal F_{(*)}}(w)$ is the formfactor describing the
hadronization  into a $D^{(*)}$  meson, and  $\cal K$  is a  known and
constant  factor.  The  Lorentz factor  $w$  of the  $c$-quark in  the
$b$-quark  rest-frame  takes  values  between  $w=1$  (``zero-recoil''
situation:  the  $c$-quark is  at  rest)  and  $w=1.5$ ($c$-quark  and
$\ell\nub$ leaving back-to-back).  In the limit of $m_Q\to \infty$ the
formfactors   ${\cal  F_{(*)}}(w)$   are  equal   to   the  Isgur-Wise
function.\cite{iwf}  Heavy quark  symmetry provides  the normalization
constraint ${\cal F_{(*)}}(1) =  1$. At zero-recoil, the light degrees
of freedom  (the spectator quark, the  sea quarks and  gluons) are not
sensitive  to the flavor  change of  the heavy  quark. Because  of the
finite mass  of the $b$ and  $c$ quarks, small corrections  need to be
computed---this  is done with  phenomenological models  or (currently
quenched) lattice QCD calculations.

The  experimental   approach  consists  in  the   measurement  of  the
differential  rate  $d\Gamma/dw$  as   a  function  of  $w$,  and  the
extrapolation of the data to  $w=1$ to obtain ${\cal F}(1)\vcb$.  This
measurement is preferentially done with \Bdstarlnu\ instead of \Bdlnu\ 
decays: The decay  rate is kinematically suppressed at  $w=1$ for both
decays,   but  less   so  for   \Bdstarlnu.   By   virtue   of  Luke's
theorem,\cite{Luke:1990eg} there are no corrections at order $1/m_{b}$
or $1/m_{c}$ for \Bdstarlnu, but they are present for \Bdlnu\ and thus
increase  the theoretical errors  in this  case.  The  background from
high-mass $X_c$ states constitutes a difficult experimental systematic
problem.  The decay \Bdlnu\ is even more affected by this, as the decay
\Bdstarlnu\  here is a  background process  with a  branching fraction
that is about two times larger.

The event selection in this analysis\cite{Aubert:2003hn} starts from a
charged  lepton ($e$ or  $\mu$) with  momentum $p^*  > 1.2\gev$  and a
reconstructed   $\Dstarp\to  \Dz\pi_s$  decay,   where  the   \Dz\  is
reconstructed   in  four   modes:  $\Dz\to   K^-\pi^+,  K_S^0\pip\pim,
K^-\pip\pim\pip,  K^-\pip\piz$.    The  mass  difference   $\deltam  =
m_{\Dz\pi_s}  -  m_{\Dz}$  is  used  for  the  selection  of  \Dstarp\ 
candidates  and the  determination  of the  combinatorial background.  
Given  the  very  large data  sample,  it  is  possible to  study  and
constrain  most of  the backgrounds  directly  in data:  We study  the
uncorrelated background (where the  lepton and \Dstarp\ originate from
different $B$  mesons) in control  samples based on the  opening angle
between  the  \Dstarp\  and  the  lepton (signal  decays  tend  to  be
back-to-back).   Continuum  background  is  reduced with  event  shape
variables,  the remaining component  is subtracted  with off-resonance
data.   The  determination  of  the  most  dangerous  background  from
high-mass $X_c$ states in data  is described in the next section.  The
only  component  taken  from  Monte  Carlo  (MC)  simulations  is  the
correlated background, where $\B\to \Dstarp X, X\to Y\ell$.

At the \FourS, the known momentum magnitude of the $B$ mesons provides
sensitivity to the missing mass  in the signal decay \Bzdstarlnu\ from
the observed  particles \Dstarp\ and  $\ell$.  Assuming that  the only
unmeasured  particle is  a massless  neutrino, the  angle  between the
momenta of the \Bz\ and the combined $\Dstar\ell$ is

\begin{equation}
\cos\theta_{B,  \Dstar\ell} =
\frac{2E_BE_{\Dstar\ell}               -               m_B^2              -
m^2_{\Dstar\ell}}{2 |\vec{p}_B||\vec{p}_{\Dstar\ell}|}.    
\label{e:dstarlnu}
\end{equation}

\begin{figure}[!ht]
 \begin{center}
  \unitlength1.0cm 
  \begin{picture}(25.,6.2)
   \put( -0.2,  0.0){\includegraphics[width=0.51\textwidth, bb= 0 150 570 690, clip=]{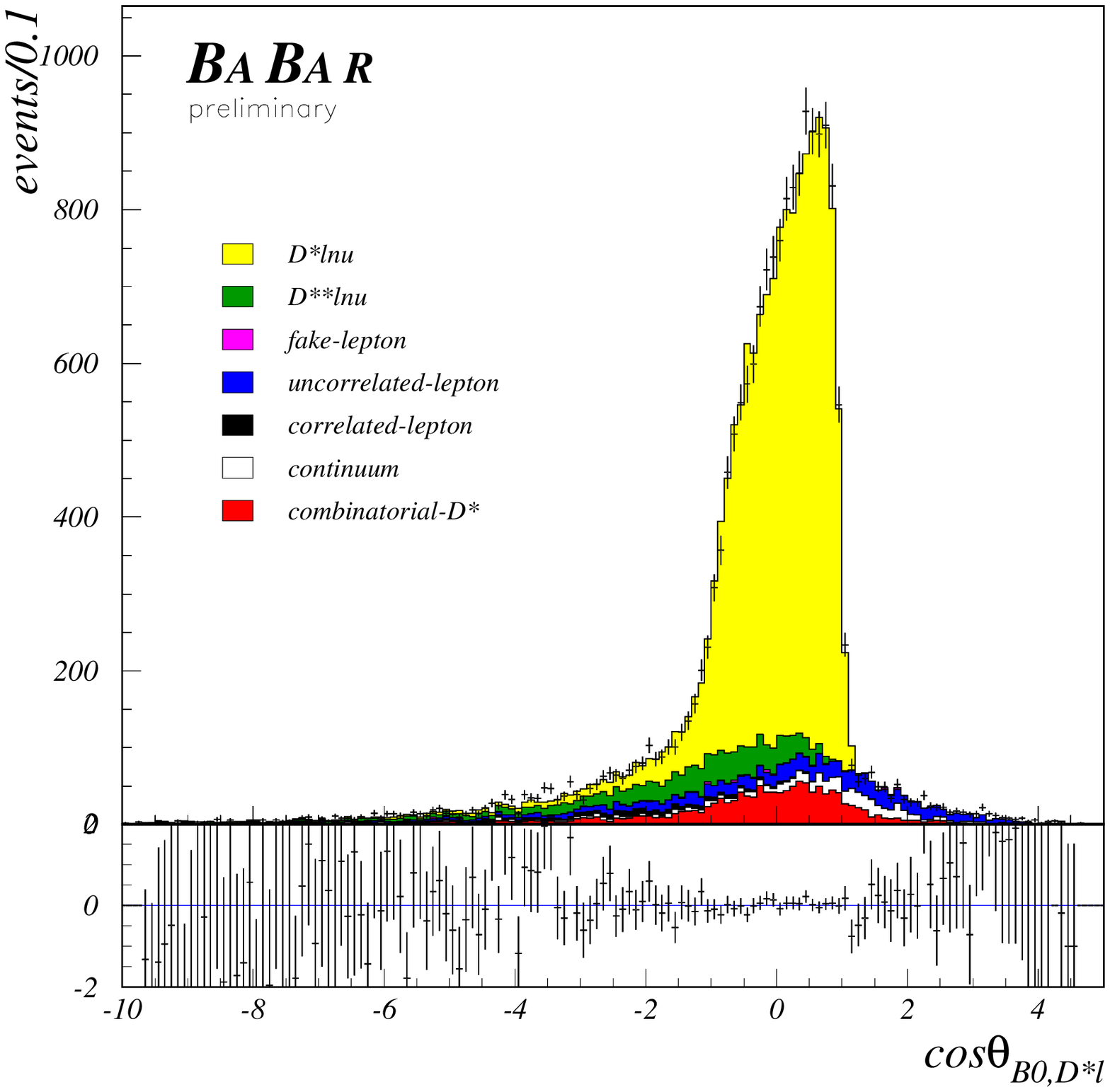}}
   \put(  7.0, -0.7){\includegraphics[width=0.47\textwidth]{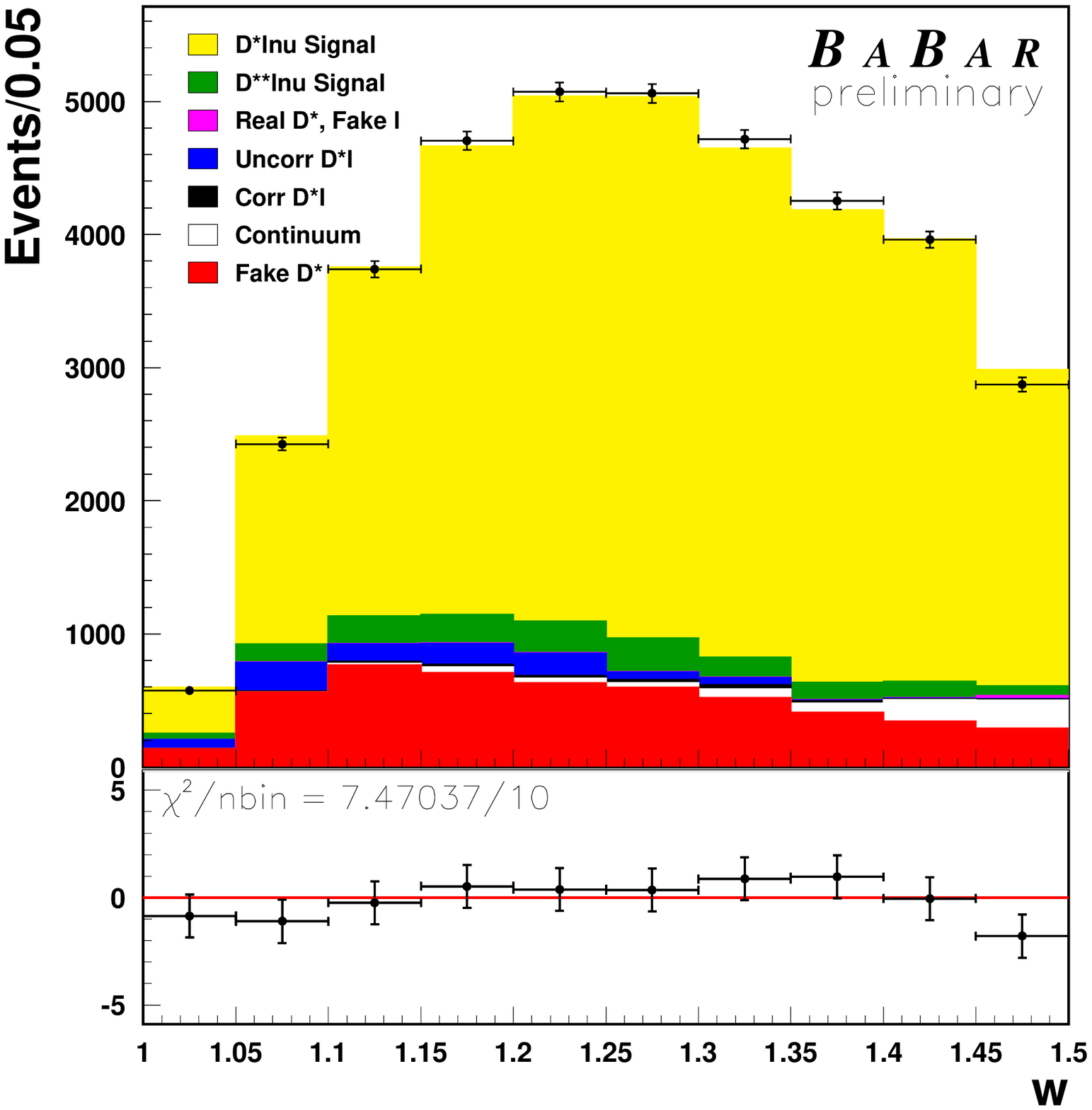}}
   \put(  5.2,  4.5){{\large\bf a)}}
   \put( 12.0,  4.5){{\large\bf b)}}
  \end{picture}
  \caption{a)  Distribution of
    $\cos\TBY$   for  the  decay   \mbox{$\bar{B^0}  \to   D^{*+}  e^-
      \bar{\nu_e}, \Dz\to  K^-\pi^+$}, where the points  show the data
    and  the filled histograms  represent the  MC components.   The MC
    normalization is determined  in a fit.  The bottom  plot shows the
    fractional  deviation  of  the   data  from  the  fit  result.   b)
    Comparison of  the $w$ distribution for $\bar{B^0}  \to D^{*+} e^-
    \bar{\nu_e}$ in data and the  result of the fit. The fit residuals
    are shown in the bottom plot. }
  \label{f:dstarlnu}
 \end{center}
\end{figure}

\noindent 
This  quantity will  lie  in the  physical  range for  a signal  decay
(modulo  resolution  effects;  signal  events  are  required  to  have
$|\cos\theta_{B, \Dstar\ell}| < 1.2$).  Decays with additional missing
particles  will  lead  to  a   tail  at  low  values,  illustrated  in
Fig.~\ref{f:dstarlnu}  for \Bzdstarenu\  decays, where  the background
from $D^{**}\ell\nub$ exhibits a long  tail.  This is also visible for
signal  decays, due to  missed photons  from bremsstrahlung.   For the
extrapolation  to  zero-recoil,  the   decay  rate  must  be  measured
differentially    in    $w    =    (m^2_{\Bz}   +    m^2_{\Dstar}    -
q^2)/(2m_{\Bz}m_{\Dstar})$,  where $q^2 =  (p_{\Bz} -  p_{\Dstar})^2$. 
The   direction    of   the    \Bz\   momentum   is    obtained   from
eq.~\ref{e:dstarlnu} up to an  azimuthal ambiguity about the direction
of  the  $\Dstar\ell$  pair:  an  unbiased estimator  of  $w$  with  a
resolution of $\sigma(w) \sim 0.02$  is calculated from the average of
the two  solutions corresponding to minimal and  maximal angle between
the \Bz\ and \Dstarp\ mesons.

The signal yield  as a function of $w$ is determined  with a fit based
on   a  quadratic  formfactor   parametrization\cite{dstarlnuff}  (see
Fig.~\ref{f:dstarlnu}).  Based on ca.~$57000$ signal events, we obtain
$\vcb      =      (37.27\pm      0.26_{stat}      \pm      1.43_{syst}
{{}^{+1.48}_{-1.23}}_{theo})\times  10^{-3}$. The  dominant systematic
errors of this  result are due to tracking and  vertexing (but not the
slow pion efficiency), \Dz\ branching fractions and $B$ lifetimes, and
$f_{00}\equiv  \cbf(\FourS\to \BzBzb)$. The  theoretical error  in the
lattice  calculation  for   $F_*(1)  =  0.913{}^{+0.030}_{-0.035}$  is
balanced  among  statistical, fitting,  matching,  spacing, mass,  and
quenching components.\cite{Hashimoto:2001nb}

The     branching     fraction     $\cbf(\Bdstarlnu)    =     (4.68\pm
0.03_{stat}\pm0.29_{syst})\%$   is  determined   by   integrating  the
differential  $w$ distribution  (cf.  eq.~\ref{e:diff})  thus reducing
the uncertainties from formfactor parametrizations.  These results are
somewhat lower  than other measurements (especially the  recent one by
CLEO\cite{Briere:2002ew}).
As  all  measurements  are  systematically  limited,  a  complementary
approach, \eg, in  the recoil of a fully  reconstructed $B$ candidate,
will be a very interesting  result addressing the largest errors.  The
theoretical uncertainties  associated with the  extrapolation to $w=1$
might be  much reduced  with a  sample of the  order of  $10^6$ signal
decays.


\section{Inclusive Cabibbo-favored Decays}

In the  description of exclusive \Bzdstarlnu\ decays,  the HQET relies
on  both $m_b$  and $m_c$  being  large.  For  inclusive decays,  this
requirement  can  be  relaxed  to  $m_b  \gg  \lqcd$,  expressing  the
separation between  the very  short time scale  relevant for  the weak
$b$-quark decay and  the long time scale for  the hadronization of the
hadronic  remnant.  The  OPE\cite{ope} can  be combined  with  HQET to
calculate, \eg, the total  semileptonic width $\Gamma_{sl}$ in a power
series  in   $\Lambda_{QCD}/m_b$  and  a   perturbative  expansion  in
$\alpha_s(m_b)$.  Incalculable quantities are parametrized in terms of
nonperturbative hadronic  matrix elements.   At lowest order,  the OPE
expression reduces to the parton model. There are no power corrections
at order $\lqcd/m_b$  in the total rate.  The  leading corrections are
parametrized with
\begin{itemize}
  \item $\lbar = m_B - m_b + (\lone + 3\ltwo)/2m_b$, the energy of the
 light degrees of freedom.  To first order, \lbar\ is
 the mass difference of the $b$ quark and the $B$ meson.
  \item \lone\ is the negative kinetic energy squared of the $b$ quark
 in the $B$ meson.
  \item \ltwo\ describes the  chromomagnetic coupling of the $b$ quark
 spin to the light degrees of freedom.
\end{itemize}


\noindent Note  that these parameters  are scheme and order  dependent. 
At higher  orders many more  parameters ($\rho_1, \rho_2,  {\cal T}_1,
\ldots {\cal  T}_4$, etc.)   enter, none of  which are  currently well
known. They constitute a large fraction of the theoretical errors.

These parameters  are not restricted  to the description of  the total
semileptonic rate, they also appear in the calculation of differential
semileptonic $B$ decay spectra and  in other $B$ decays: \lbar\ can be
related to the mean photon energy of the decay \bsg, and \ltwo\ can be
determined from the mass difference of $B^*$ and $B$ mesons.

The  parameters  can be  related  to the  shape  of  decay spectra  in
semileptonic  \Bxlnu\  decays,  such  as the  lepton  energy  spectrum
$E_\ell$ or  the invariant hadronic mass spectrum  \mx.  The inclusive
description  of  semileptonic  \Bxlnu\  decays  does  not  distinguish
between specific hadronic  final states $X$, \eg, the  $D$ and \Dstar\ 
mesons.   To compare, \eg,  the theoretical  expectation for  the \mx\ 
distribution with the measured  spectrum, it is therefore necessary to
resort to  observables smearing  the differential spectrum.   A simple
and sensitive  possibility is given  by moments of various  order.  As
different  moments have  different dependencies  on the  parameters, a
simultaneous  fit  to several  moments  provides  for an  experimental
determination of the  nonperturbative parameters.  The large branching
fraction for \Bxlnu\ allows for  a very precise determination of these
parameters.   We   can  therefore  shift  a  large   fraction  of  the
theoretical errors into (smaller)  experimental errors, which are more
amenable to  a proper  statistical interpretation.  This  is essential
for a  quantitative understanding of  the errors of the  extracted CKM
parameters.

\subsection{Inclusive Semileptonic Branching Fraction}
The model-independent measurement  of the total inclusive semileptonic
branching   fraction  $\cbf(\Bxenu)$  was   pioneered  by   the  ARGUS
collaboration.\cite{Albrecht:1993pu}  It  has a small model-dependence
in the sense that no assumptions  on the shape of the primary electron
spectrum    from   \Bxenu\    decays   are    necessary.     In   this
analyis,\cite{Aubert:2002uf} we  select events  by the presence  of a
high-momentum tag  electron ($1.4 < p^*  < 2.3\gev$).  In  the rest of
the  event,  signal electrons  are  identified  and  grouped into  two
separate  classes,  depending on  whether  they  have opposite  charge
(``unlike  sign sample'':  the  two electrons  are  either from  $B\to
X_{\bar{c}} e^+ \nu,  \Bb\to X_c e^- \nu$ or  from $\Bb\to X_c e^-\nu,
X_c \to  Y e^+ \nu$) or  the same charge (``like  sign sample'': $B\to
X_{\bar{c}}  e^+ \nu,  \Bb\to B\to  X_{\bar{c}}  e^+ \nu$  and $B  \to
X_{\bar{c}} e^+ \nu, \Bb \to X_{\bar{c}}, X_{\bar{c}} \to Y e^+\nu$)
as the tag electron.

\begin{figure}[!ht]
 \begin{center}
  \unitlength1.0cm 
  \begin{picture}(25.,4.0)
   \put( -0.1,  0.0){\includegraphics[width=0.49\textwidth]{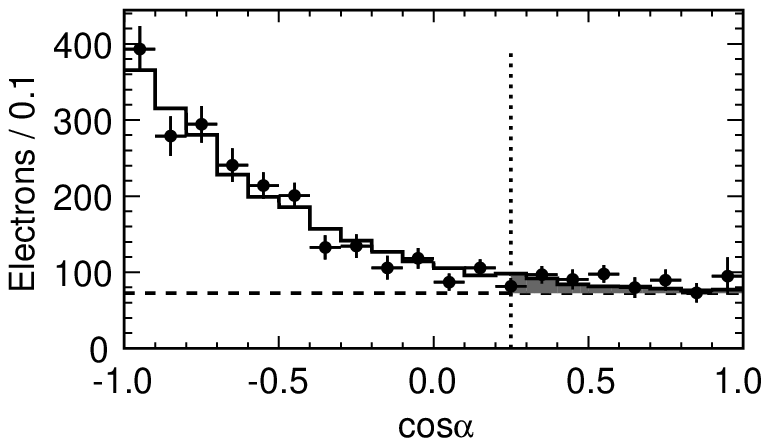}}
   \put(  6.3,  0.1){\includegraphics[width=0.51\textwidth]{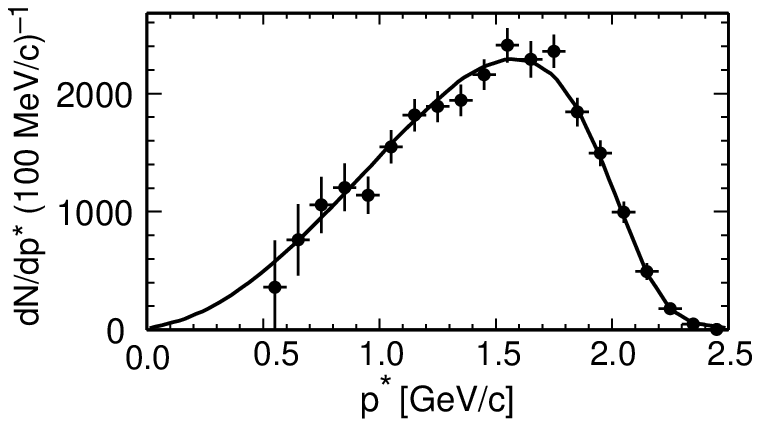}}
   \put(  5.3,  3.0){{\large\bf a)}}
   \put( 11.9,  3.0){{\large\bf b)}}
  \end{picture}
  \caption{a) Cosine of the opening angle between the signal
    electron  with $0.7 <  p^* <  0.8\gevc$ and  the tag electron  in the
    unlike-sign sample. The shaded area represents the background
    electrons, the vertical line illustrates the minimum requirement on the
    opening angle. b) Momentum distribution of electrons from \Bxenu\
    decays, after efficiency and bremsstrahlung corrections. }
  \label{f:islbf}
 \end{center}
\end{figure}

In the unlike-sign class, the background $\Bb\to X_c e^-\nu, X_c \to Y
e^+ \nu$ can be strongly  reduced by exploiting the fact that electron
pairs from  the same $B$ are preferentially  back-to-back, while there
is no  correlation in the opening  angle distribution $\alpha(e_{tag},
e_{signal})$ for  the case of the  pair coming from  two different $B$
mesons.  This is illustrated in Fig.~\ref{f:islbf}a, where the opening
angle distribution  for signal  electrons with $0.7  < p^*  < 0.8\gev$
shows  a flat signal  component and  a background  contribution rising
toward low values.  The shape information is taken from MC simulation,
and  the  data distribution  is  fitted  to  determine the  background
contribution  (shaded) in  the signal  region.  The  requirement  of a
small opening  angle also  removes most of  the heavy  pair background
(from $J/\psi\to  e^+e^-$; the  remaining pair background  from photon
conversions and Dalitz  $\piz\to e^+e^-\gamma$ decays is reconstructed
and removed explicitly). The effect of \BzBzb\ mixing can be unfolded,
as
\begin{eqnarray}
 {1\over \varepsilon_{\alpha}(p^*)}{dN_{\pm\mp}\over dp^*} &=&
 {dN_{primary}\over dp^*}\cdot(1 - \chi) + {dN_{casc}\over dp^*}\cdot\chi \\ 
 {dN_{\pm\pm}\over dp^*} &=&
 {dN_{primary}\over dp^*}\cdot\chi + {dN_{casc}\over dp^*}\cdot(1 - \chi),
\end{eqnarray}

\noindent where $\chi = f_{00}\cdot\chi_d = 0.087$ (here $\chi_d = 0.174\pm0.009$ is the
\BzBzb\ mixing  parameter\cite{Groom:in} and $f_{00} =  0.50$ has been
assumed)  and the  efficiency $\varepsilon_{\alpha}(p^*)$  of  the opening
angle requirement.  

The integration of  the spectrum over the range $0.6  < p^* < 2.5\gev$
(see  Fig.~\ref{f:islbf}b) yields  $N(\Bxenu)  = 25070\pm  410_{stat}$
signal events  for an integrated luminosity of  about $4\invfb$.  Only
small  corrections  need  to  be applied:  Bremsstrahlung  corrections
(2.2\%), geometric  acceptance (16\%), event selection  bias (2\%) and
the extrapolation  (6.1\%) to $p^* =  0$.  From these  numbers and the
overall normalization  from the number  of tag electrons  we determine
$\cbf(\Bxenu) = (10.87\pm 0.18  \pm 0.30)\%$.  The dominant systematic
error  arise  from  electron  identification plus  tracking  and  from
semileptonic  decays  of   upper-vertex  charm  particles,  especially
affected  by the poor  knowledge of  $\cbf(D_s\to\phi\pi)$.  Extending
the measurement  range to lower values  of $p^*$ does  not improve the
error,  as backgrounds  (pair background,  cascade decays)  with large
uncertainties grow very large.

From the total branching fraction, \vcb\ can be extracted, \eg, in the
$1S$  expansion.\cite{Hoang:1998ng} The  errors in  this  approach are
dominated  by  theoretical  uncertainties.   In the  next  section  an
alternative  method   is  described  that  takes   into  account  more
information   and  their   correlations.    There,  the   differential
measurement  of  the  lepton  energy  spectrum and  its  moments  will
contribute substantially.

The measurement  of $\Bxenu$  has also  been done in  the recoil  of a
\breco\  candidate,\cite{eps2001} albeit  with much  less  statistics. 
This allows a comparison of electron energy spectra from \Bz\ and \Bp\ 
decays and provides  one way to study effects  of quark-hadron duality
violation and other nonperturbative effects like weak annihilation and
Pauli interference.

\subsection{Hadronic Mass Moments}
In \Bxlnu\  decays, the invariant  mass \mx\ distribution is  the most
sensitive probe  to physics beyond the  parton model and  hence to the
nonperturbative  parameters  \lbar\  and  \lone.   The  lepton  energy
distribution  also  has sensitivity,  but  at  a  reduced level.   The
experimental  feasibility matches  this situation:  the  lepton energy
distribution can be obtained with high precision and resolution from a
measurement of  the electron only, a relatively  straightforward task. 
On the other  hand, the reconstruction of the  complete hadronic final
state is a much more involved procedure.


The  event selection in  this analysis\cite{Aubert:2003dr}  requires a
\breco\ candidate and an identified  lepton with $p^* > 900 \mevc$ and
charge consistent  for a primary  $B$ decay.  The charge  imbalance of
the event is  required to be not larger than  one. These criteria lead
to a data sample of about 7100 events.

All remaining charged tracks and  neutral showers that are not part of
the \breco\  candidate are combined  into the hadronic system  $X$.  A
neutrino  candidate is  reconstructed from  the  missing four-momentum
$p_{miss}  = p_{\FourS}  - p_X  - p_{\breco}$,  where all  momenta are
measured  in  the  laboratory  frame.   Consistency  of  the  measured
$p_{miss}$  with   the  neutrino  hypothesis  is   enforced  with  the
requirements $E_{miss} > 0.5\gev$,  $|\vec{p}_{miss}| > 0.5 \gev$, and
$|E_{miss} -  |\vec{p}_{miss}|| < 0.5\gev$.  The  determination of the
mass  of the  hadronic  system is  improved  by a  kinematic fit  that
imposes four-momentum conservation, the  equality of the masses of the
two  $B$ mesons,  and forces  $p_{\nu}^2  = 0$.   The resulting  $\mX$
resolution is $350\mevcc$ on  average.  MC simulated event samples are
used to calibrate the absolute mass scale, determine efficiencies, and
estimate backgrounds.   This mass scale calibration  allows the direct
determination of the moments of the \mx\ distribution without recourse
to simulated decay spectra.

\begin{figure}[!ht]
 \begin{center}
  \unitlength1.0cm 
  \begin{picture}(25.,7.)
   \put(-0.2, -1.2){\includegraphics[width=0.49\textwidth]{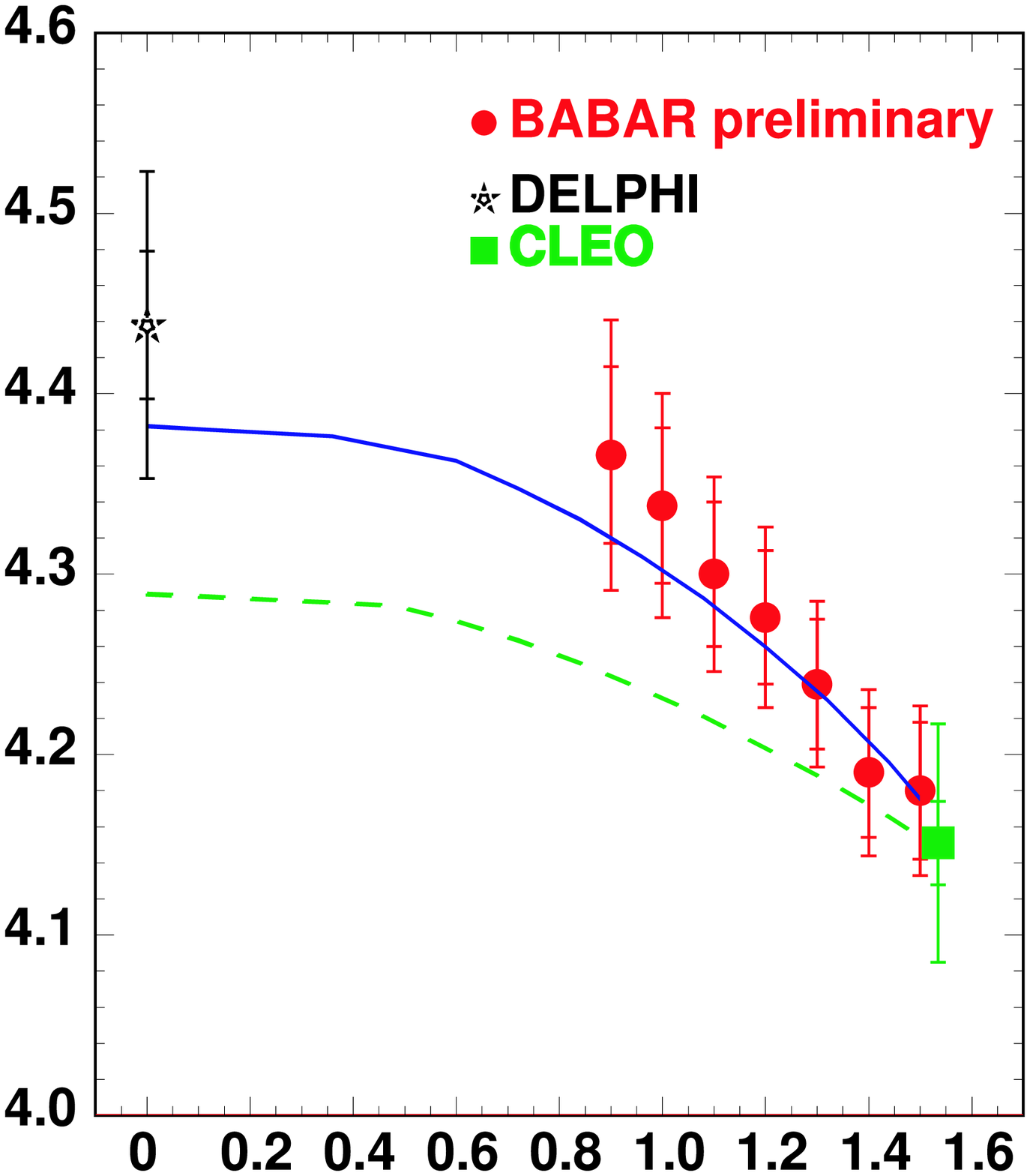}}
   \put( 6.1, -0.0){\includegraphics[width=0.585\textwidth]{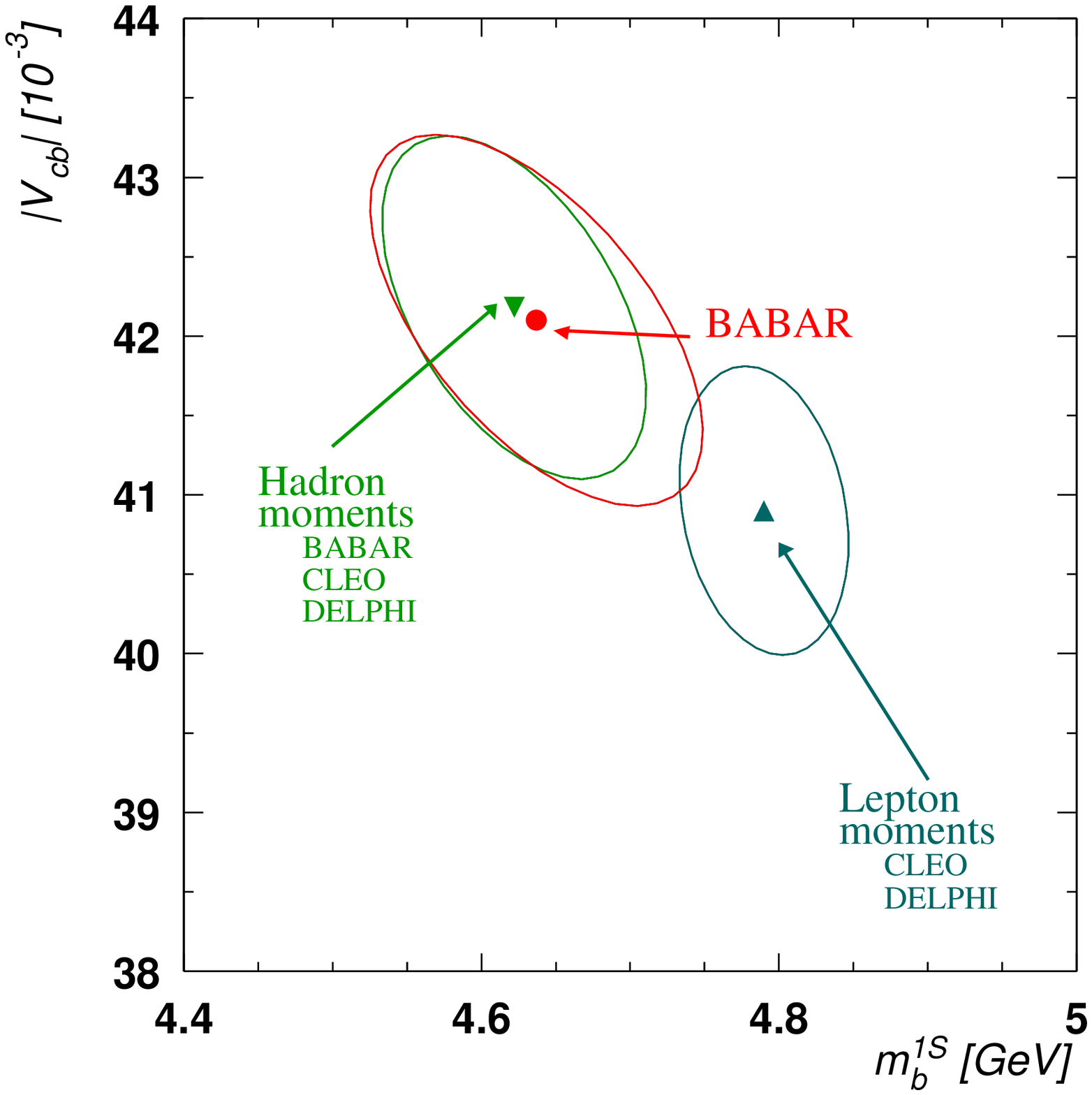}}
   \put( 4.2, 0.1){{\large $p^*_{min} [\gev]$}}
   \put(-0.1, 4.4){\rotatebox{90}{\large $\langle m_X^2\rangle [\gev^2]$}}
   \put( 1.0, 5.7){\includegraphics[height=0.8cm]{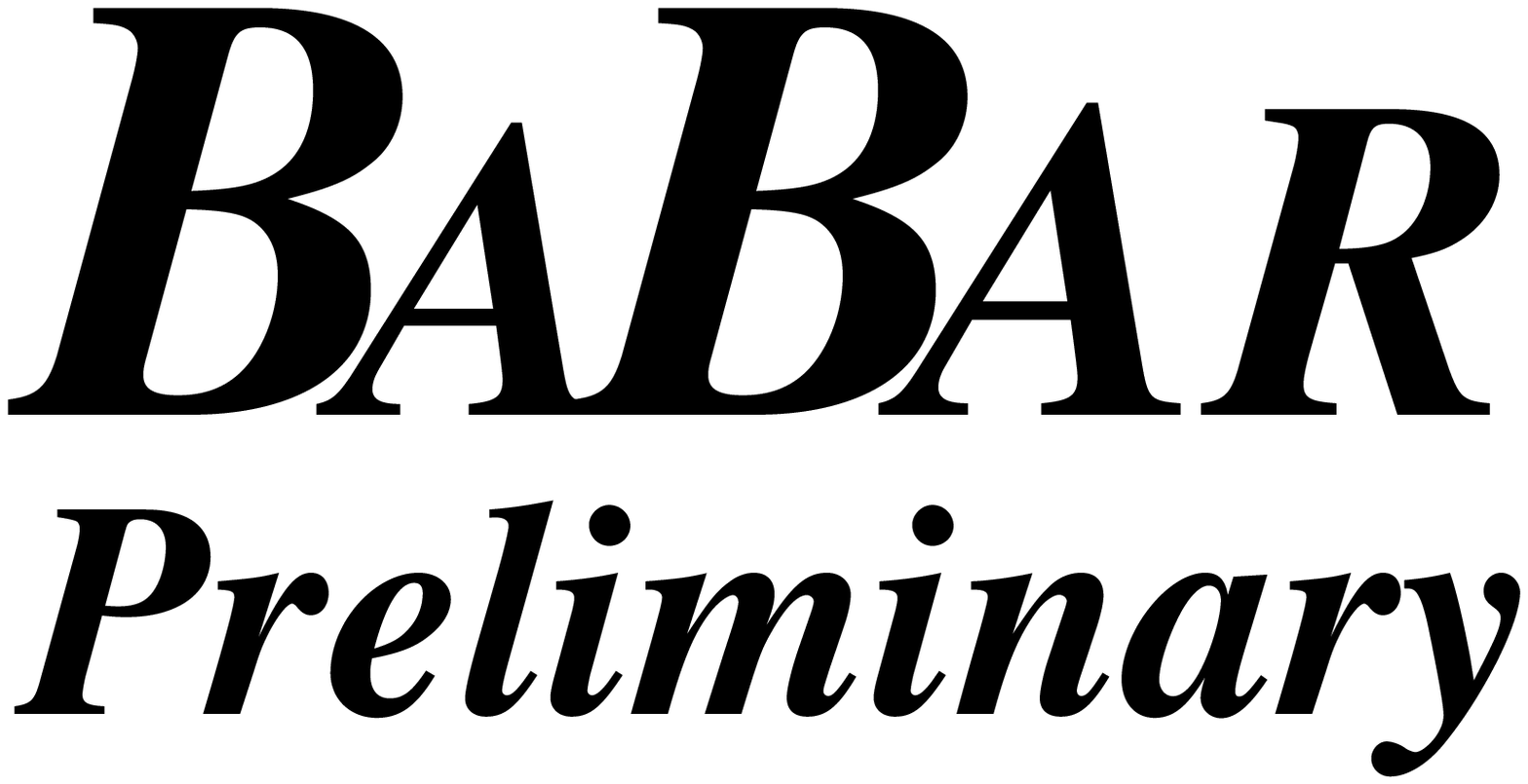}}
   \put(10.7, 5.6){\includegraphics[height=0.8cm]{babarPrel.eps}}
   \put( 1.2, 1.2){{\large \bf a)}}
   \put( 7.5, 1.2){{\large \bf b)}}
  \end{picture}
  \caption{(a) Measured hadronic mass moments for different lepton threshold
    momenta  $p^*_{min}$.   The   errors  of  the  individual  \babar\ 
    measurements   are  highly   correlated.    For  comparison,   the
    measurements by the DELPHI and CLEO collaborations are also shown.
    The solid curve is a fit  to the \babar\ data; the dashed curve is
    the OPE prediction based on  the CLEO result combined with information
    from the  decay \bsg.  (b) Constraints  on the $b$  quark mass and
    \vcb\ from  this measurement,  and the fit  to the  combined hadron
    moments and lepton moments, respectively.}
  \label{f:moments}
 \end{center}
\end{figure}

The resulting  moments of  the hadronic mass-squared  distribution are
shown as  a function of  the threshold lepton momentum  $p^*_{min}$ in
Fig.~\ref{f:moments}a.  A substantial rise of the moments toward lower
momentum is  visible, due to  the enhanced contributions  of high-mass
charm states (phase-space suppressed  at higher $p^*_{min}$). The main
contributions to  the systematic error are from  the detector response
simulation  and   from  semileptonic  decays   of  upper-vertex  charm
particles.  The  uncertainty from the  modeling of the $X_c$  state is
negligible compared to the other systematic errors.

Accounting  for  all correlations  between  the  moments at  different
$p^*_{min}$,  we  obtain  $\lbar   =  0.53\pm0.09\gev$  and  $\lone  =
-0.36\pm0.09\gev^2$      in       the      \MSb\      regularization
scheme.\cite{Falk:1997jq}   The   errors    given   do   not   include
uncertainties due  to terms  at ${\cal O}(1/m_B^3)$.   For comparison,
Fig.~\ref{f:moments}a  also  shows the  result  of  the hadronic  mass
measurement of  DELPHI,\cite{Calvi:2002wc} fully consistent  with this
result.   The CLEO  result\cite{Cronin-Hennessy:2001fk}  of the  first
hadronic mass moment at $p^*_{min}  = 1.5\gev$ is also consistent, but
in   combination   with   the    mean   photon   energy   from   $b\to
s\gamma$\cite{Chen:2001fj}  shows a  different  $p^*_{min}$ dependence
(taking into  account the bias  from the limited photon  energy range,
the agreement is much better\cite{Bigi:2003zg}).

A fit  to all hadronic moments  from \babar\ is performed  in the $1S$
scheme,\cite{Bauer:2002sh} as this scheme exhibits better convergence
of the  perturbative series than  other alternatives. The  results are
$m_b^{1S}  =  4.638  \pm  0.094_{exp} \pm  0.062_{dim\oplus  BLM}  \pm
0.065_{1/m_B^3}   \gev$   and  $\lone   =   -0.26\pm  0.06_{exp}   \pm
0.04_{dim\oplus  BLM}  \pm   0.04_{1/m_B^3}  \gev^2$.   The  fit  also
utilizes    the   semileptonic    width   $\Gamma_{sl}    =   (4.37\pm
0.18)\times10^{-11}\mev$ (determined  from \babar\ data)  to determine
$\vcb   =  (42.10   \pm  1.04_{exp}   \pm  0.52_{dim\oplus   BLM}  \pm
0.50_{1/m_B^3})\times 10^{-3}$.

The consistency of  the OPE is tested by  combining the measurement of
\babar\  with the  four lepton  energy  moments measured  by the  CLEO
collaboration\cite{Briere:2002hw}         and        the        DELPHI
collaboration.\cite{Calvi:2002wc}  In  Fig.~\ref{f:moments}b, the  fit
results  are  shown  separately  for  hadron mass  and  lepton  energy
moments. The $\Delta\chi^2  = 1$ contours of hadronic  mass and lepton
energy moments do  not overlap. Note that the  theoretical errors here
do not include uncertainties due  to terms at ${\cal O}(1/m_B^3)$.  In
the  future,  the  extension  of  these measurement  to  include  more
high-precision  observables will  allow  a quantification  as to  what
precision can be expected realistically from the OPE.

\section{Inclusive Cabibbo-suppressed Decays}
In  the measurement  of  \Bxulnu\ decays,  the  large background  from
\Bxclnu\ decays  has to be reduced  by restricting the  phase space in
the  analyses.   One  possibility  is  to measure  the  lepton  energy
spectrum at the ``endpoint'', beyond the kinematic cutoff for \Bxclnu\
decays, at  $p^* > 2.3\gev$.  A  disadvantage of this  approach is that
only  about 10\% of  all charmless  semileptonic decays  are measured.
This   leads  to  a   significant  extrapolation   with  corresponding
uncertainties.  The model-dependence of this error can be reduced with
information  on the movement  of the  $b$ quark  inside the  $B$ meson
obtained    from    the    photon    energy    spectrum    in    \bsg\
decays.\cite{Bornheim:2002du}

Alternative methods have been proposed,  such as the invariant mass of
the hadronic system X in \Bxlnu\ decays.\cite{Barger:tz} Here 50--80\%
of  all \Bxulnu\  decays  are  measured, depending  which  cut $\mx  <
\mxcut$ is used for the determination  of the signal yield.  As in the
case of the endpoint spectrum, there is a dependence on the light-cone
distribution function (``shape function'') of the $b$ quark in the $B$
meson, describing  the Fermi motion of  the heavy quark  in the meson.
Another process  sensitive to  this shape function  is the  rare decay
\bsg,  with  a   branching  fraction\cite{Chen:2001fj}  $\cbf(\bsg)  =
3.21\pm0.53\times   10^{-4}$.   This   rate  is   even   smaller  than
$\cbf(\Bxulnu)$, thus  limiting the precision of  the determination of
the  shape  function  parametrization  and parameters.   If  the  HQET
parameters determined  in Cabibbo-favored semileptonic  $B$ decays can
be  used   consistently  in   the  shape  function   description,  the
determination  of   \vub\  will   benefit  from  the   high  precision
measurements described in the previous section.

The total rate  can be translated to \vub\ with an  error of about 5\%
from uncertainties of higher  orders in the perturbative expansion and
the uncertainty of the $b$ quark mass.\cite{Hoang:1998ng,Uraltsev:1999rr}

\subsection{Endpoint Spectrum}
\begin{figure}[!ht]
 \begin{center}
  \unitlength1.0cm 
  \begin{picture}(25.,6.)
   \put(  0.0,  0.0){\includegraphics[width=0.49\textwidth, bb= 0 60 410 445, clip=]{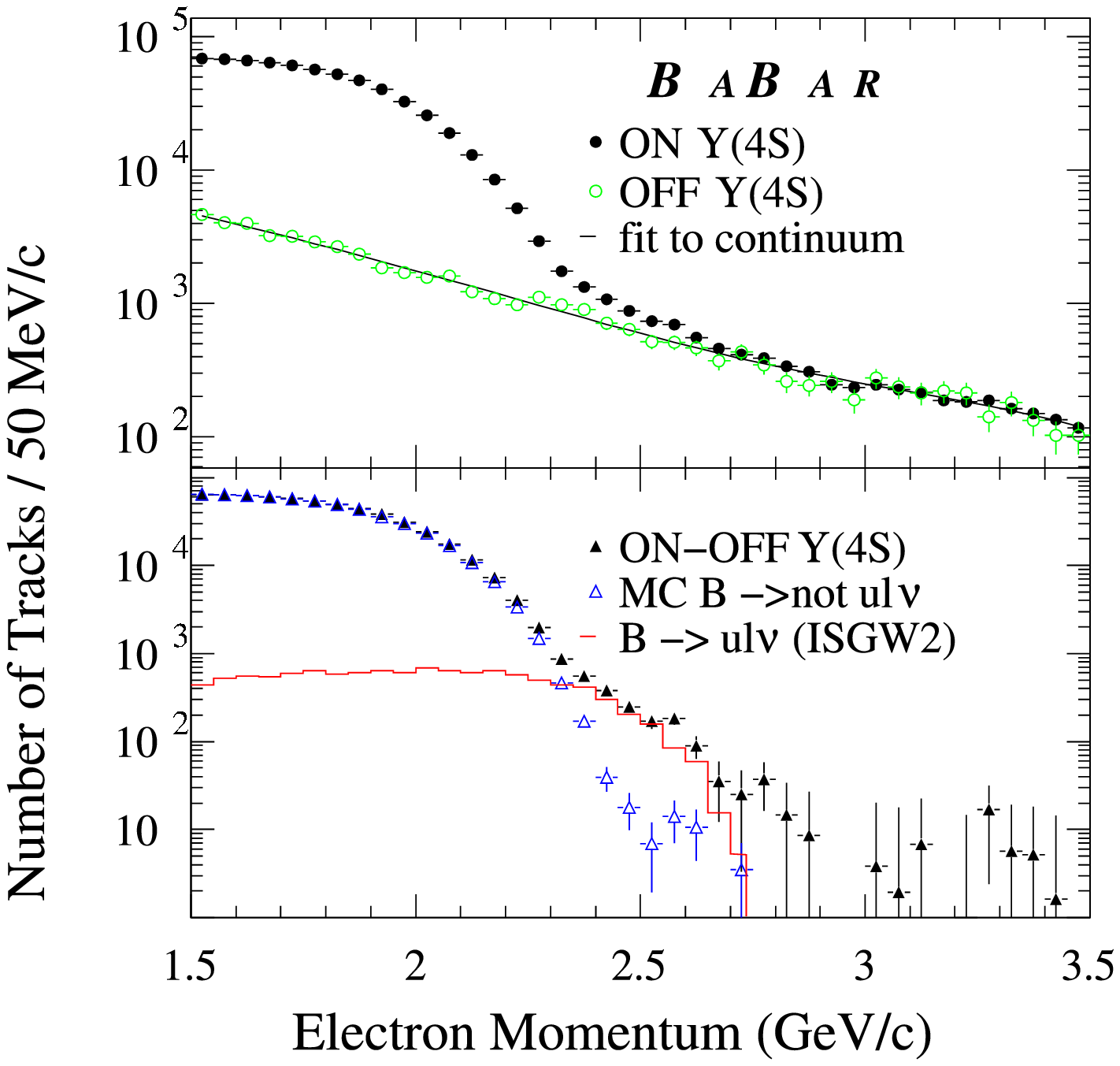}}
   \put(  6.8,  0.0){\includegraphics[width=0.49\textwidth]{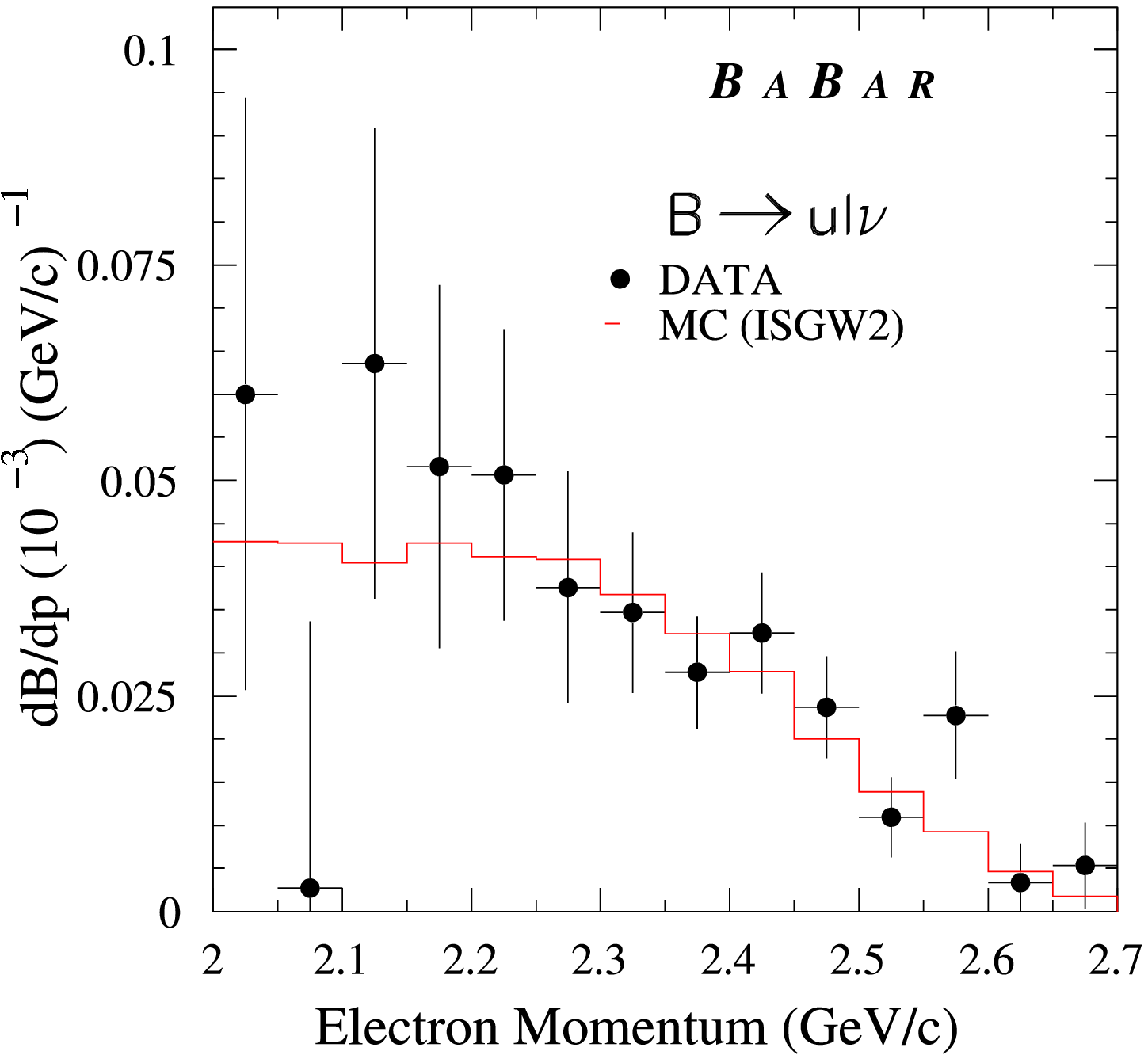}}
   \put(  5.4,  5.3){{\large\bf a)}}
   \put(  5.4,  2.7){{\large\bf b)}}
   \put( 12.3,  5.2){{\large\bf c)}}
  \end{picture}
  \caption{a) Electron momentum distribution in the \FourS\
    rest-frame for on-resonance and off-resonance data, respectively.
    b)  Electron momentum  distribution  after continuum  subtraction,
    with signal  and background MC distributions.  c) The differential
    branching fraction  as a function of the  electron momentum, after
    efficiency and  bremsstrahlung corrections. The  data are compared
    to the prediction  of the ISGW2 model (assuming  a total inclusive
    branching fraction of $10^{-3}$ for \Bxulnu\ decays with mass $\mx
    < 1.5\gevcc$).}
  \label{f:endpoint}
 \end{center}
\end{figure}

In this analysis,\cite{Aubert:2002pi} the  event selection is based on
a  high-momentum electron  $(p^* >  2.0\gev)$ and  the signature  of a
neutrino. Specifically, we require  for the missing momentum $p_{miss}
>1\gev$  to be  pointing into  the  main detector  acceptance $-0.9  <
\cos\theta^*_{miss}  < 0.8$  and  in the  hemisphere  opposite to  the
electron.   At  low  momenta,  the  electron sample  is  dominated  by
electrons  from \Bxclnu\  decays  over a  continuum background.   This
latter component dominates  the spectrum at high momenta.   We fit the
continuum background  with a  $4^{th}$ degree Chebyshev  polynomial in
the off-resonance data and the high-momentum range in the on-resonance
data.  After subtraction,  the signal events are visible  in the range
$2.3 < p^*  < 2.6\gev$ and illustrated with  the solid (red) histogram
in  Fig.~\ref{f:endpoint}b.  The  restriction to  this  momentum range
yields  a  total of  $1696\pm  133$ signal  events  with  a signal  to
background ratio  of $S/B  = 0.25$.  Extending  the momentum  range to
lower  values decreases  $S/B$ due  to more  background  from \Bxclnu\ 
decays,  leading  to   substantially  higher  uncertainties  from  the
modeling of \Bxclnu\ decays.  On  the other hand, the extrapolation is
decreased, resulting  in a smaller theoretical error.  In the endpoint
range,  the partial  branching fraction  is determined  to  be $\Delta
\cbf(\Bxulnu)  =  (0.152 \pm  0.014_{stat}  \pm 0.0014_{syst})  \times
10^{-3}$, where  the dominant errors  arise from the  uncertainties in
the continuum subtraction, the motion  of the $B$ meson in the \FourS\ 
rest-frame, and the selection efficiency.

From  this  result,  the   extrapolation  to  the  total  semileptonic
charmless   branching    fraction   is    done   as   in    the   CLEO
analysis.\cite{Bornheim:2002du} Here the shape function parameters are
determined by a fit to the \bsg\ photon energy spectrum. The result is
$\cbf(\Bxulnu)  = (2.05\pm  0.27_{exp}\pm  0.46_{f_u})\times 10^{-3}$,
where  the errors are  now grouped  into a  first part  containing the
statistical and  systematic uncertainty from  the endpoint measurement
and a second part describing the uncertainties from the extrapolation.
This            yields           $\vub=            (4.43           \pm
0.29_{exp}\pm0.50_{f_u}\pm0.35_{s\gamma}\pm0.25_{\Gamma})\times 10^{-3}$.

\subsection{Hadronic Mass Spectrum}
In this analysis\cite{Aubert:2003zw} we use the invariant mass \mx\ of
the  hadronic system  to separate  \Bxulnu\ decays  from  the dominant
\Bxclnu\  background  in  events  tagged by  the  fully  reconstructed
hadronic  decay  of  a   \breco\  candidate.   This  method  offers  a
substantially larger signal acceptance  than the endpoint measurement. 
The hadronic  system $X$  in the decay  \Bxlnu\ is  reconstructed from
charged  tracks   and  energy  depositions  in   the  calorimeter  not
associated with the \breco\ candidate or the identified lepton.
We require  exactly one  charged lepton with  $p^* > 1  \gevc$, charge
conservation ($Q_{X}  + Q_\ell + Q_{\breco}  = 0$), and  $\mmiss < 0.5
\gev^2$.   
We reduce  the $\Bzb\to\Dstarp\ell^-\overline{\nu}$ background  with a
partial  reconstruction of  the decay  (using the  $\pi^+_s$  from the
$\Dstarp\to \Dz\pi_s^+$  decay and the lepton).   Furthermore, we veto
events with charged or neutral kaons in the recoil \Bb.

In  order  to reduce  experimental  systematic  errors (in  particular
lepton identification), we determine  the ratio of branching fractions
\rusl\ from  $N_u$, the observed  number of $\Bxulnu$  candidates with
$\mx<1.55$\gevcc,  and $N_{sl}  = 29982\pm233$,  the number  of events
with at least one charged lepton:
\begin{displaymath}
\rusl=
\frac{\BR(\Bxulnu)}{\BR(\Bxlnu)}=
\frac{N_u/(\varepsilon_{sel}^u \varepsilon_{\mx}^u)}{N_{sl}} 
\times \frac{\varepsilon_l^{sl} \varepsilon_{reco}^{sl} } {\varepsilon_l^u \varepsilon_{reco}^u }.
\label{eq:vubExtr}
\end{displaymath}
Here $\varepsilon^u_{sel} = \allepsu\pm0.006_{stat}$ is the efficiency
for  selecting  \Bxulnu\ decays  once  a  \Bxlnu\  candidate has  been
identified,  $\varepsilon^u_{\mx} =  \allepsmx\pm0.009_{stat}$  is the
fraction of  signal events with the reconstructed  $m_X < 1.55\gevcc$;
$\varepsilon_l^{sl}/\varepsilon_l^u  =  0.887\pm0.008_{stat}$ corrects
for the  difference in the efficiency  of the lepton  momentum cut for
\Bxlnu\           and          \Bxulnu\           decays,          and
$\varepsilon_{reco}^{sl}/\varepsilon_{reco}^u =  1.00 \pm 0.03_{stat}$
accounts  for  a  possible  efficiency difference  in  the  $B_{reco}$
reconstruction in events with \Bxlnu\ and \Bxulnu\ decays.

\begin{figure}[!ht]
 \begin{center}
  \unitlength1.0cm 
  \begin{picture}(25.,6.)
   \put(  0.0,  0.0){\includegraphics[width=0.49\textwidth]{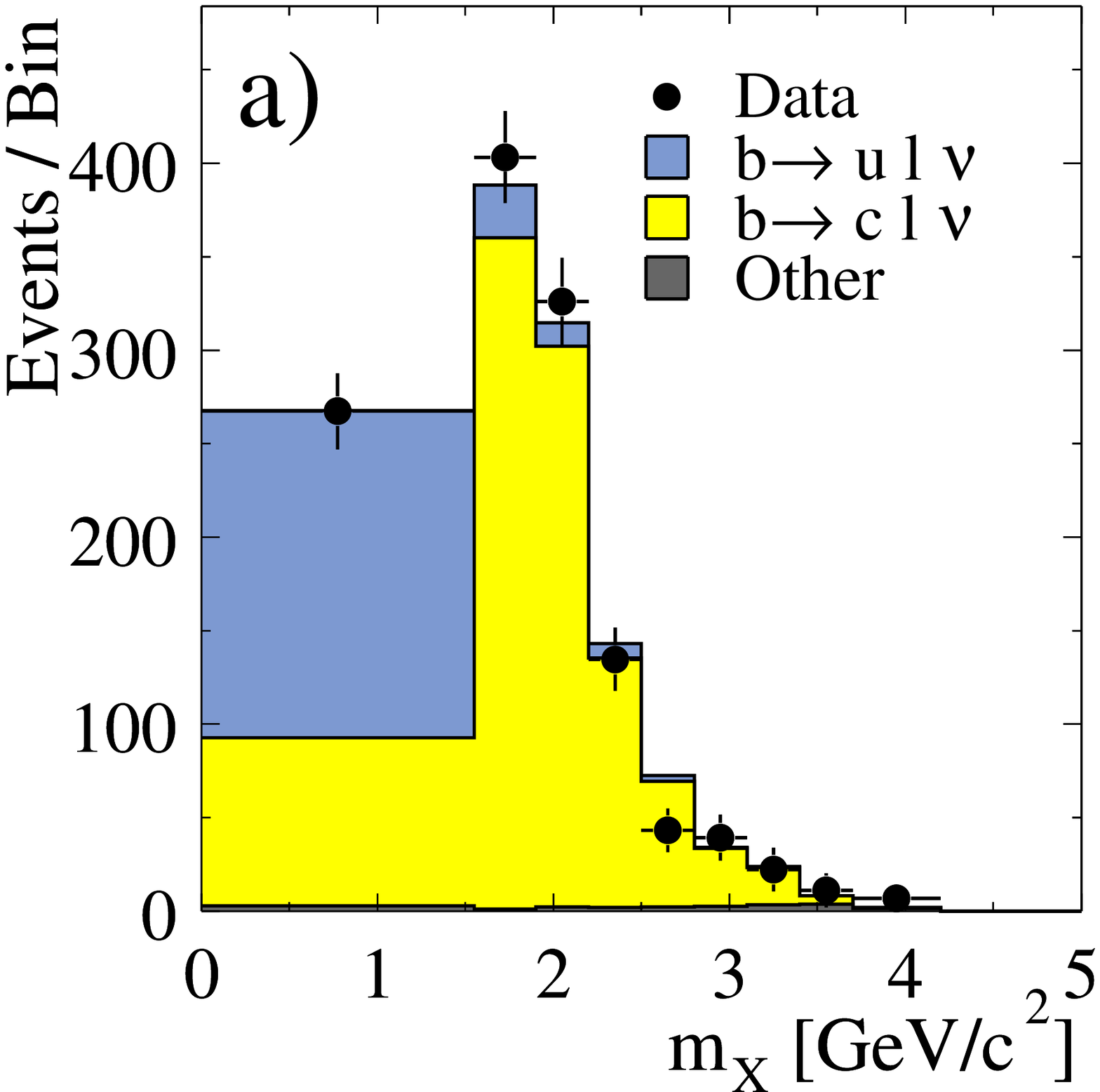}}
   \put(  6.8,  0.0){\includegraphics[width=0.49\textwidth]{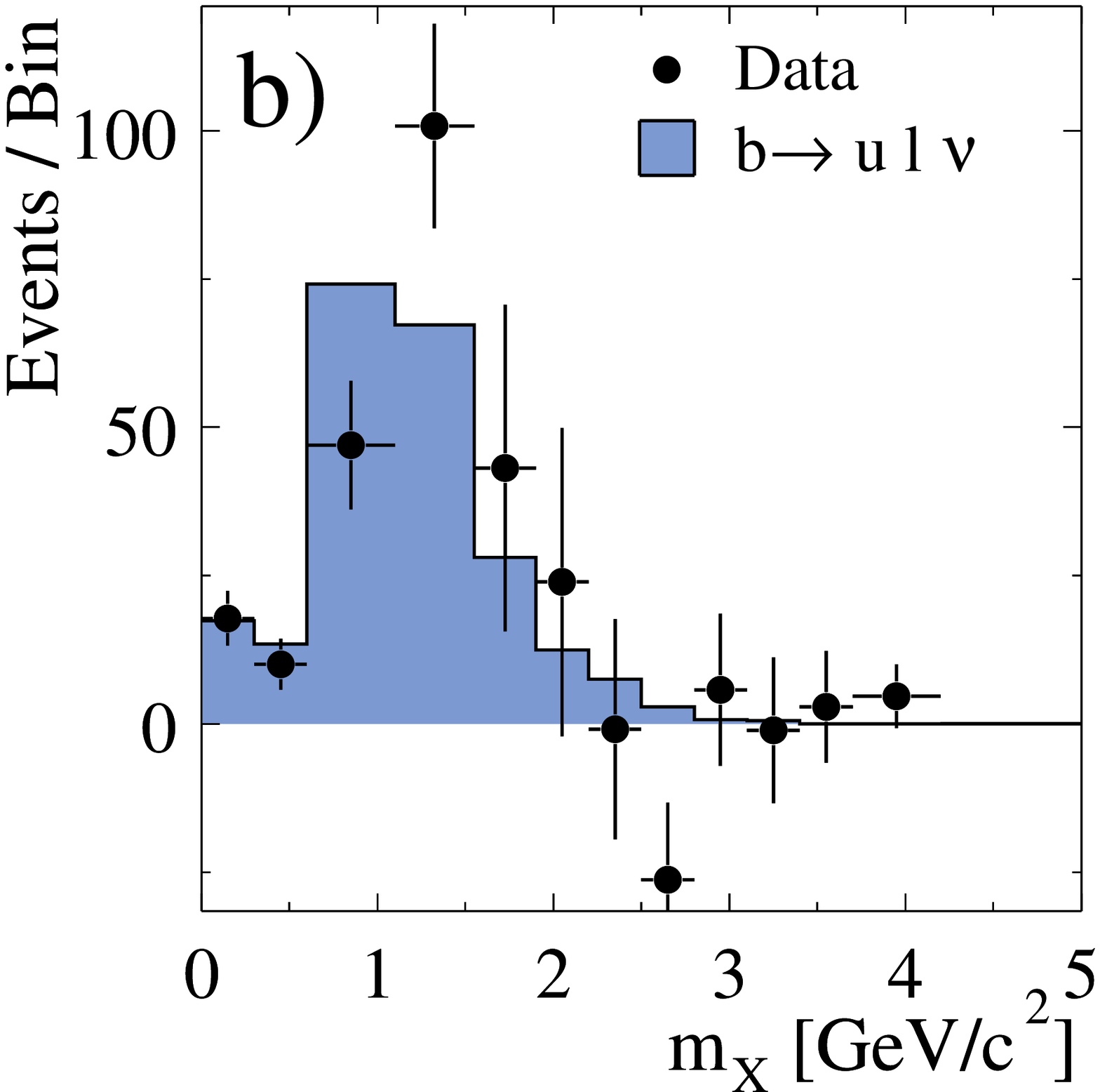}}
  \end{picture}
  \caption{The $\mx$ distribution for \Bxlnu\ candidates: a)
    data (points) and fit components, and b) data and signal MC after
    subtraction of the $b\to c\ell\nu$ and the ``other'' backgrounds.}
  \label{f:mxhad}
 \end{center}
\end{figure}

We extract  $N_u$ from the $\mx$ distribution  by a fit to  the sum of
three contributions:  signal, background  $N_{c}$ from \Bxclnu,  and a
background  of $<1\%$ from  other sources.   In each  bin of  the \mx\ 
distribution, the combinatorial  $B_{reco}$ background for $\mes>5.27$
is  subtracted on  the basis  of  a fit  to the  $\mes$ distribution.  
Fig.~\ref{f:mxhad}a shows the  fitted $\mx$ distribution.  To minimize
the model dependence, the first bin is extended to $\mx < 1.55\gevcc$.
We find
$175\pm 21$ signal events and $90\pm5$ background events in the region
$\mx < 1.55\gev$. From this we determine
$\rusl = (\allbrbrVal \pm \allbrbrEcstat_{stat}
\pm \allbrbrEsyst_{syst}
\pm\allbrbrEthHi_{theo})\times 10^{-2}.   $ The dominant  detector systematic
errors   are  due  to   the  uncertainty   in  photon   detection  and
combinatorial     background     subtraction.     The     efficiencies
$\varepsilon_{sel}^{u}$  and $\varepsilon_{\mx}^{u}$ are  sensitive to
the modeling  of the \Bxulnu\  decays.\cite{DeFazio:1999sv} We assess
the   theoretical  uncertainties   by   varying  the   nonperturbative
parameters  within their  errors, $\lbar  =  0.48 \pm  0.12 \gev$  and
$\lone  =   -0.30\pm0.11  \gev^2$,   obtained  from  the   results  in
Ref.~\refcite{Cronin-Hennessy:2001fk}  by removing  terms proportional
to $1/m_b^3$  and $\alpha_s^2$ from the relation  between the measured
observables and \lbar\ and \lone.   Here we assume that the parameters
of the shape function are given by the HQET parameters.

Combining the  ratio \rusl\  with the measured  inclusive semileptonic
branching fraction,\cite{Aubert:2002uf} we obtain
$\BR(\Bxulnu) = 
(\allbr
\pm\allbrE
\pm\allbrEsyst
\pm\allbrEthHi)\times10^{-3}$. 
With  the   average   $B$  lifetime\cite{Hagiwara:fs} we obtain
$
\vub = (\allvub
\pm\allvubE_{stat}
\pm\allvubEsyst_{syst}
\pm\allvubEthHi_{theo} \pm\allvubEtheo_{\Gamma})\times 10^{-3}$, where
the last error is the uncertainty in the extraction of \Vub\ from the
total decay rate. No error is
assigned to the assumption of parton-hadron duality.  This result is
consistent with previous inclusive measurements, but has a smaller
systematic error, primarily due to larger acceptance and higher sample
purity.

\section{Exclusive Cabibbo-suppressed Decays}

The  reconstruction  of   exclusive  charmless  semileptonic  \Bxulnu\
decays, where $X_u = \pi, \rho, \omega, \ldots$, is challenging due to
large   backgrounds  and  the   missing  neutrino.    The  theoretical
description  is  on  less  solid  foundations than  in  the  inclusive
case. The determination of \vub\ in this case requires the computation
of formfactors, parametrizing the  behavior of the hadronic current in
the $B$  meson decay.   Due to  the light mass  of the  hadronic final
state, HQET is of much less help here compared to \Bdstarlnu, and most
of the  calculations are based on  phenomenological models. Currently,
lattice   QCD   calculations  are   still   based   on  the   quenched
approximation, with uncertainties of the order of 15--20\%.  The decay
\Bpilnu\  is  more  amenable  to  these calculations  than  the  decay
\Brholnu, where  the hadronic final  state is a broad  resonance.  The
computation  of formfactors  is only  possible  in the  region of  low
momentum  $p_\pi \leq  1\gev$.   This corresponds  to  high $Q^2  \geq
17\gev^2$, where  the rate is kinematically  suppressed.  In addition,
this is precisely the  kinematic region where the current experimental
methods  are   limited  by  very  high  backgrounds.   With  the  high
luminosities  achievable  at the  $B$  factories,  the measurement  of
exclusive decays  in the  recoil of a  \breco\ candidate  offers large
advantages.

\subsection{\Bzrhoenu}
This   analysis\cite{Aubert:2003zd}   aims   for   a   high   neutrino
reconstruction  efficiency   and  is   similar  to  a   previous  CLEO
analysis.\cite{Behrens:1999vv}    Events   are   selected    with   a
high-momentum  electron and  divided  into two  samples  based on  the
electron momentum:  The high-momentum $high$-$E_e$  with $2.3 <  E_e <
2.7\gev$  and $low$-$E_e$  $2.0 <  E_e <  2.3\gev$.   The $high$-$E_e$
sample is  primarily used for  the determination of the  signal, while
the  $low$-$E_e$ sample  serves for  the measurement  of  the \Bxclnu\ 
background. The analysis is optimized for the measurement of
$B\to\rho e^+\nu$, but also selects $B\to\pi e^+\nu$ and $B^+\to\omega
e^+\nu$ to better control cross-feed background contributions. 

Continuum  background  is  suppressed  with  a neural  network  of  14
variables.  As the hadronic final  state in this exclusive analysis is
much less constrained than  in \Bdstarlnu, the neutrino reconstruction
carries  more  weight.   The  direction  of the  missing  momentum  is
required to point into the main detector $(|\cos\theta_{miss}| < 0.9)$
to reject  events with  substantial energy loss  along the beam  axis. 
The angle $\alpha$ between  the reconstructed missing momentum and the
inferred  neutrino momentum  is required  to be  small  $(\cos\alpha >
0.8)$ and (as in the decay \Bdstarlnu) the angle between the $B$ meson
momentum  and  the combined  $eh$  momentum  (where  $h =  \pi,  \rho,
\omega$) is  required to lie in the  physical region $(|\cos\theta_{B,
  eh}| < 1.1)$.

\begin{figure}[!ht]
 \begin{center}
  \unitlength1.0cm 
  \begin{picture}(25.,6.)
   \put(  0.0,  0.0){\includegraphics[width=0.49\textwidth]{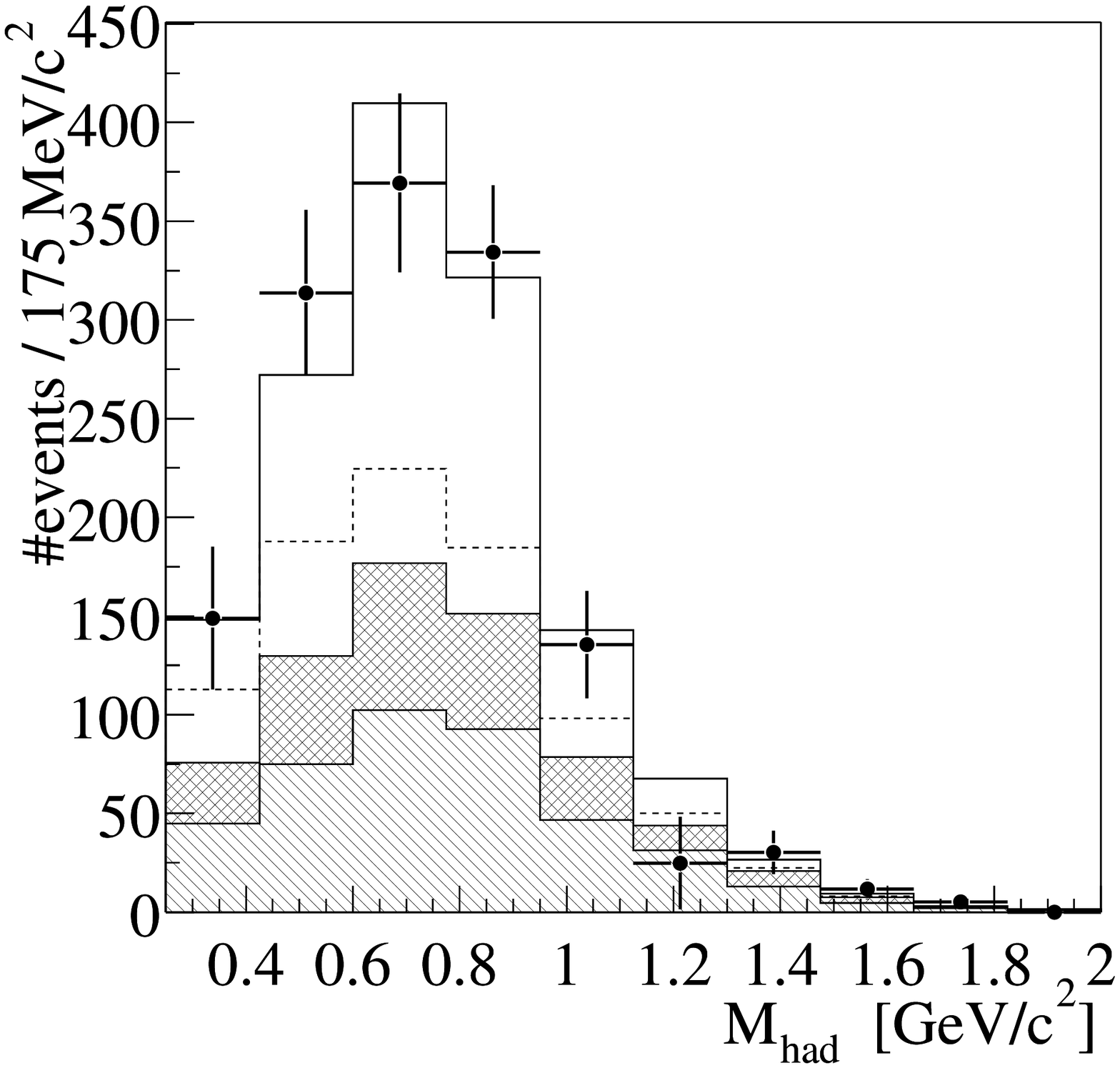}}
   \put(  6.8,  0.0){\includegraphics[width=0.49\textwidth]{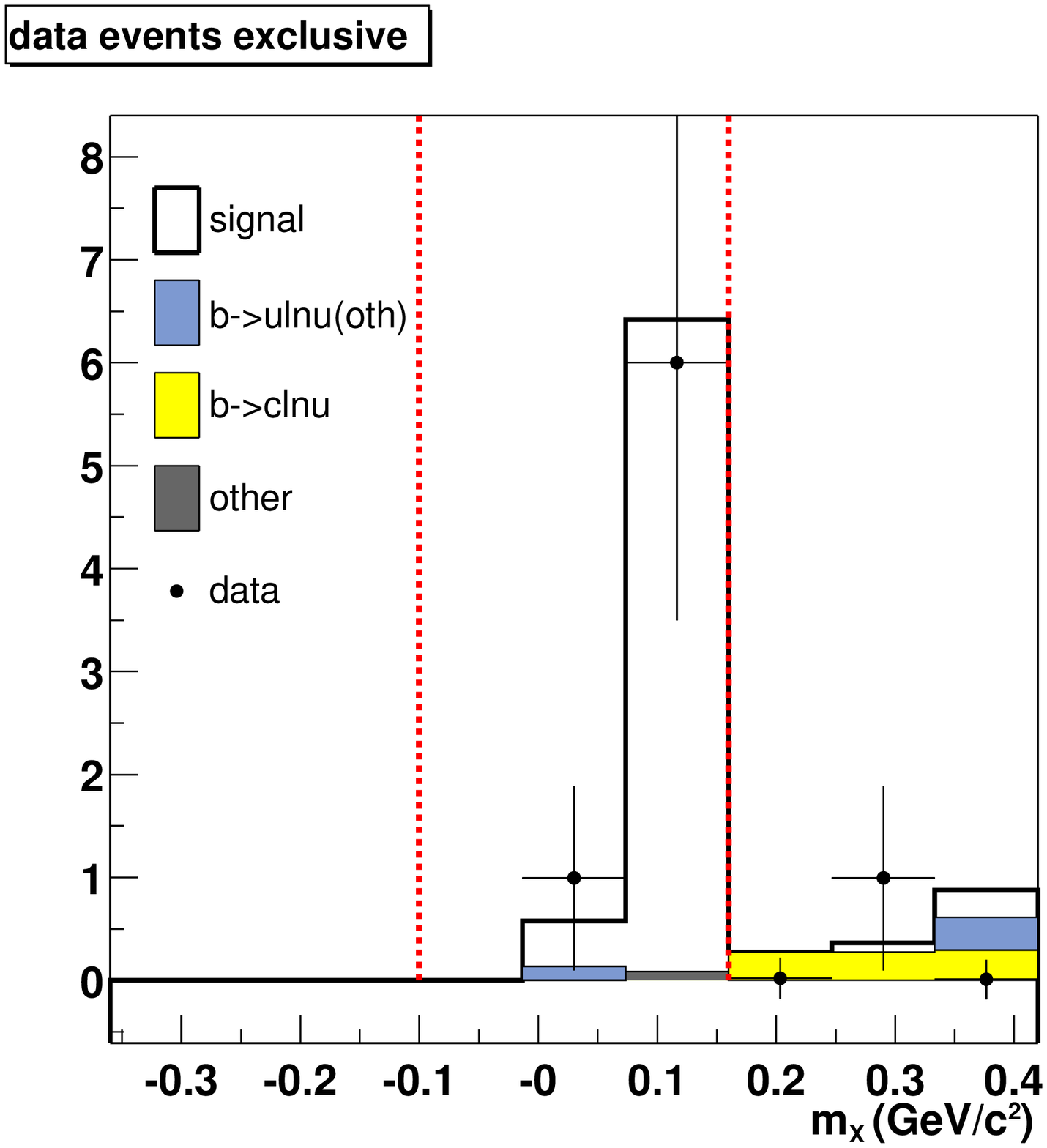}}
   \put(  4.7,  5.0){{\large\bf a)}}
   \put( 11.5,  5.0){{\large\bf b)}}
  \end{picture}
  \caption{Illustration  of the signal to background ratio for exclusive
    charmless semileptonic decays.  a) Continuum-subtracted fit
    projections for $m_{\pi^0\pi^-}$ for the $\Bz\to \rho^- e^+\nu$
    channel in the $high$-$E_e$ electron-energy region.  The
    contributions are the direct and crossfeed components of the
    signal (unhatched region, above and below the dashed line,
    respectively), the background from $b\to u e \nu$ (double-hatched
    region), and $b\to c e \nu$ and other backgrounds (single-hatched
    region).  b) The \mx\ projection for \Bmpilnu\ measured on the
    recoil of a fully reconstructed hadronic $B$ decay.}
  \label{f:rhoenu}
 \end{center}
\end{figure}

The signal  yield is extracted by  a binned maximum  likelihood fit to
$high$-$E_e$ and $low$-$E_e$ samples  in two variables: $\de = E_{had}
+ E_e  + p_{\nub} - E_{beam}$  (where $p_{\nub}$ is  obtained from the
missing  momentum) and  the mass  of  the hadronic  system $m_{had}  =
m_{\pi\pi(\pi)}$. The  shape of  the continuum distributions  is taken
from off-peak data, the remaining  shapes are from MC simulations. The
fit    incorporates    isospin   and    quark    model   relations.    
Fig.~\ref{f:rhoenu}a  illustrates  the  $m_{\pi\pi}$ variable  in  the
$high$-$E_e$  sample.   For  an   integrated  luminosity  of  $\clu  =
50\invfb$,  a  signal  yield  of  $S =  505\pm  63_{stat}$  events  is
obtained,  leading  to  a  branching fraction  of  $\cbf(\Bzrhoenu)  =
(3.29\pm  0.42_{stat}\pm 0.47_{syst}\pm  0.60_{theo})\times  10^{-4}$. 
The CKM matrix  element \vub\ is determined from  the relation $\vub =
\sqrt{\frac{\cbf(\Bzrhoenu)}{\Gamma_{theo}\tau_{\Bz}}}   =  (3.64  \pm
0.22_{stat}    \pm    0.25_{syst}   {{}^{+0.39}_{-0.56}}_{theo})\times
10^{-3}$.  The  dominating systematic error  are from the  modeling of
resonant and nonresonant \Bxulnu\ decays, the tracking efficiency, and
the fit  method. The  theoretical error is  determined as half  of the
full  spread  of  all  theoretical  uncertainties  in  the  formfactor
calculations.\cite{Scora:1995ty}
A measurement of  the $Q^2$ dependence will help  to reject models not
describing  \Brhoenu\  decays,  but  will  not help  in  reducing  the
inherent model dependence of  the error. Here, only unquenched lattice
QCD calculation of the formfactors and the experimental measurement in
the same kinematic region will advance the field.

\subsection{Exclusive Decays on the Recoil}
This analysis\cite{delRe:xslb2u}  is a combination  of the high-purity
event tagging  based on a  fully reconstructed hadronic $B$  decay and
the  exclusive reconstruction  of signal  decay in  the  recoil.  This
approach results in a very  low overall signal efficiency of the order
of  0.1\%, but  allows the  measurement  of \Bpilnu\  over the  entire
kinematic range.  This type of  measurement will become a prime method
for the determination of  \vub\ from exclusive semileptonic decays, as
the traditional approach mentioned in the previous section is affected
by very low $S/B$ problems,  especially in the range where lattice QCD
can  provide model-independent  formfactor calculations.   Because the
statistical yield  of the method  is low, an integrated  luminosity of
about $\clu \sim 500\invfb$ is needed before the method will provide a
better measurement of \Bpilnu\ than the traditional approach.

While all exclusive decays can  be measured in this analysis paradigm,
only the  measurement \Bpilnu\ shall be described  here (for \Bmpilnu)
as  it is the  most promising  channel for  lattice QCD  calculations. 
After the requirement of a  fully reconstructed hadronic $B$ decay for
tagging purposes, events with  semileptonic $B$ decays are selected by
the  requirement  of a  muon  or electron  with  $p_\ell  > 1.0\gev$.  
Cabibbo-favored \Bxclnu\  decays are  rejected by requirements  on the
missing mass $m_{miss}^2 < 0.4\gev^2$, the invariant mass of the \piz\ 
candidate and the requirement that  no other hadronic charged track be
detected. The  number of signal  $S = 7.0\pm2.6$ events  is determined
from a fit to the \mes\ distribution of selected events, corrected for
background  $B =  0.2\pm  0.2$ (determined  from  MC simulations)  and
selection efficiency $\varepsilon  = 0.42\pm0.04$ for all requirements
after the \breco\ and lepton  candidate selections.  As in the case of
the inclusive \Bxulnu\ analysis, the signal yield is normalized to the
number of  events with a charged  lepton in the recoil  of the \breco\ 
candidate.      The     result     of     this     analysis     yields
$\cbf(\Bmpilnu)/\cbf(\Bxlnu)  = (0.76 \pm0.31_{stat}  \pm 0.11_{syst})
\times  10^{-3}$.   This result  is  statistics  limited; the  largest
systematic  errors are  uncertainties of  the \mes\  fits,  a possible
selection  bias  for charmless  semileptonic  $B$  decays compared  to
general semileptonic $B$ decays, and the measurement of photons in the
calorimeter.


\section{Conclusions}
In  the last  few  years, the  study  of semileptonic  $B$ decays  has
offered   many  new   perspectives.   Theoretical   uncertainties  are
parametrized  in  terms  of  experimental  observables,  substantially
reducing the model-dependent component in the total error on \vcb. The
large luminosity at experiments  like \babar\ at the \FourS\ resonance
allows to select  events by means of fully  reconstructed hadronic $B$
decays and  studying the  semileptonic decay of  the other  $B$ meson.
This opens the  precise study of spectral moments  in semileptonic $B$
decays  and therefore  the precision  determination of  \vcb\  and the
underlying fundamental  parameters of the theory.  It  also allows for
better  constraints in  the  study  of \Bxulnu\  decays  and leads  to
improved determinations  of \vub.  In  the future, the  measurement of
exclusive  charmless  semileptonic  $B$  decays  in  combination  with
unquenched lattice  QCD calculations  will challenge the  precision of
inclusive  \vub\  determinations.  In  the  far  future, leptonic  $B$
decays may provide yet another way to \vub.

\section{Acknowledgements}
It   is   a   pleasure   to   acknowledge   the   collaboration   with
O.~Buchm\"uller,  D.~del  Re,  R.~Faccini,  E.~Hill, V.~  L\"uth,  and
A.~Sarti.   I  have  enjoyed  instructive discussions  with  C.~Bauer,
T.~G.~Becher, Z.~Ligeti, and N.~Uraltsev.


\end{document}